\documentclass[preprint2]{aastex}
\newcommand{\s}{$_{\rm s}$}
\newcommand{\kms}{km~s$^{-1}$}
\newcommand{\etal}{{\it et al.}}
\newcommand{\ie}{{\it i.e.}}
\newcommand{\MK}{{{\rm M}_{\rm ext,K}-5\log h}}
\newcommand{\magarc}{{\rm mag\,arcsec$^{-2}$}}
\newcommand{\ri}{{$r_{21}$}}
\shorttitle{Extinction in Galaxies in the NIR}
\shortauthors{Masters \etal}
\usepackage{epsfig}
\begin{document}
\title{Internal Extinction in Spiral Galaxies in the Near Infrared }
\author{Karen L. Masters, Riccardo Giovanelli and Martha P. Haynes}
\affil{Center for Radiophysics and Space Research, Cornell University, Ithaca, New York 14853}
\email{masters@astro.cornell.edu, riccardo@astro.cornell.edu, haynes@astro.cornell.edu}

\begin{abstract}
In order to study the effects of internal extinction in spiral galaxies we search for correlations of near infrared (NIR) photometric parameters with inclination.
We use data from the 2 Micron All-Sky Survey (2MASS) Extended Source Catalog (XSC) on 15,224 spiral galaxies for which we also have redshifts. For 3035 of the galaxies, I-band photometry is available which is included in the analysis. 
From the simple dependence of reddening on inclination we derive a {\it lower limit} to the difference in magnitude between the face-on and edge-on aspect of 0.9, 0.3 and 0.1 magnitudes in I (0.81$\mu$m), J (1.25 $\mu$m) and H (1.65 $\mu$m) bands. 
We find that the faintest isophotal radius reported in the XSC (at the 21st \magarc ~level) is closer to the centers of the galaxies than other common isophotal measures ({\it e.g.} the 23.5 \magarc ~radius in I-band), and argue that it should not be assumed to represent an outer isophote at which galaxies are transparent at all viewing angles. A simple linear extinction law (\ie ~$\Delta M = \gamma \log(a/b)$) is not adequate for the full range of disk inclinations and we adopt both a bi-linear and a quadratic law. A simple photometric model is used to explain the observed behavior. Internal extinction depends on galaxy luminosity. We show that for galaxies with a K\s ~total magnitude dimmer than -20, -20.7 and -20.9 the data indicates zero extinction in J, H and K\s ~respectively, while disk opacity increases monotonically with increasing disk luminosity. 
\end{abstract}


\section{Introduction}
The determination of an accurate value for the internal extinction in 
galaxies is of relevance both in the understanding of their structure and in applications such as the determination of redshift independent distances. Edge-on spirals are used preferentially in studies that apply the luminosity-linewidth relation (Tully \& Fisher 1977) and these are, of course, the galaxies in which extinction effects will be the largest. Inadequate corrections for extinction will produce systematic distance errors in our catalogs. Recent discussions of this issue can be found in \citet{opacitybook} and \citet{cal01}, while a comprehensive history of the study of internal extinction in galaxies can be found in the introduction of \citet{Huzthesis}. Analyses similar to the ones carried out in this paper can be found in Giovanelli \etal ~(1994; hereafter G94, and 1995) and \citet{t98}.

Since there is intrinsic variability in the properties of galaxies, the determination of the amount of extinction is most readily obtained in a statistical way, in which case large samples are desirable. However, sample selection effects need to be well understood in order to avoid interpreting spurious effects as astrophysically significant. There will be further discussion of the effects of such biases on the results of this paper in Sections 5.3 and 6.2.

In this paper we report on statistical tests which search for the effects of extinction in spiral galaxies in the near infra-red (NIR) bands, I (0.81 $\mu$m), J (1.25 $\mu$m) , H (1.65 $\mu$m) and K$_{\rm s}$  (2.17 $\mu$m). NIR bands are often used in cases where accurate total magnitudes of galaxies are needed, as they clearly minimize, if not eliminate, the need for extinction corrections. For this reason it is particularly interesting to search for the effects of extinction in these bands. 

Samples and available data are described in Sections 2 and 3.
In Section 4 we apply simple tests for reddening with inclination and in Sections 5 and 6 we discuss in more detail the effects of extinction on the isophotal radius and total magnitudes respectively. A simple photometric model for the extinction is discussed in Section 7 and in Section 8 luminosity dependences are discussed. A summary of the conclusions is given in Section 9.

\section{Description of the Samples}

 The Two Micron All-Sky Survey (2MASS) completed its coverage of the near infra-red sky in February 2001.
Data were taken in three NIR bands, J, H and K$_{\rm s}$ using two identical 1.3m telescopes, one in Arizona, the other in Chile. With an extended source sensitivity of 14.8, 14.0 and 13.4 mag at J, H and K\s ~\citep{J00} and an angular resolution of about 2-3 arcseconds, 2MASS has detected more than 1.65 million galaxies \citep{J03}; the resultant data are available in the Extended Source Catalog (XSC)\footnote{Available through the online catalog query service of the IR Science Archive: \url{http://irsa.ipac.caltech.edu/applications/Gator/}}. 

 Two samples were extracted from the 2MASS All-Sky release XSC.
For statistical studies the samples should be as large as possible, but it is also important for them to have well defined selection criteria. The first sample (and by far the largest) consists of all galaxies with well defined spiral morphology and known redshift taken from the Arecibo General Catalog (a private compilation by the authors known as the AGC). Nearby, peculiar motions contribute significantly to observed recessional velocities so we further limit the AGC to $cz > 3000$ \kms ~in order to use the redshift as a reliable distance indicator. This large sample is referred to as the AGC$z$. The AGC is an all-sky compilation, but has varying completeness functions in different areas of the sky. 
A second, smaller sample consists of galaxies in the AGC$z$ for which I-band photometry is also available (as will be described below). This sample we call the SFI2 as it contains all of the SFI data \citep{g97, h99a, h99b} plus similar but newer and as yet unpublished additions to that. This sample is a homogeneous all-sky sample of mostly Sc galaxies. In summary:
\begin{description}
\item[AGC$z$] Spiral galaxies of known redshift from the Arecibo General Catalog. The redshift was further confined to $cz >$ 3000 \kms. This sample includes a total of 15244 galaxies.
\item[SFI2] Galaxies from AGC$z$ which also have I-band photometry available. This sample has a total of 3035 galaxies, which are mostly of types Sc and Sbc.
\end{description}

 The detection fractions of these galaxies in the 2MASS XSC (within 12\arcsec ~of the
previously cataloged position) are 80\% 
for the AGC$z$ and 85\% 
for the SFI2 sample. We expect that this is due in part to the detection limits of the 2MASS survey and in part to the accuracy of the currently tabulated positions of the galaxies in the AGC. In support of this assumption the detection fraction increases for the SFI2 galaxies that are expected to have more accurate positions in the AGC. For the Second Incremental release of 2MASS data, a discussion of completeness is included in the on-line Explanatory Supplement\footnote{The Explanatory Supplement to the 2MASS Second Incremental Release \citep{cutri} is an on-line document: \url{http://www.ipac.caltech.edu/2mass/releases/docs.html}}. When the search for a match is extended to a radius of 0.5\arcmin ~of known galaxies from the Zwicky Catalog, the detection rate is 95\%. 
In the NIR, surface brightness decreases towards later type galaxies \citep{J03} making late type, gas rich galaxies (like the Sc galaxies which make up the bulk of the SFI2 sample) harder to detect than E/S0/Sa galaxies.

\section{Description of data}

 For all of the galaxies in our samples various data are available. These included redshifts, coarse blue magnitudes and angular sizes from the AGC, photometry from 2MASS and I-band photometry for the SFI2 sample. In the computation of absolute magnitudes a value of H$_\circ = 100 h$ \kms ~Mpc$^{-1}$ is adopted.

\subsection{I-band photometry}

The I-band photometry available for the SFI2 sample is described in \citet{h99a}. Quantities of particular interest are:
\begin{description}
\item[$(a/b)_{\rm I}$] - I-band axial ratio derived from a fit of elliptical surface brightness contours to the I-band images. The average error in $\log(a/b)$ is 0.03.

\item[$r_{23.5}$] - radius measured at the I-band isophote of 23.5 \magarc.
\item[$m_{\rm I}$] - total I-band magnitude. Calculated from an extrapolation of a fit to the disk to 8 scale lengths of the disk. Typical total errors are in the range 0.03-0.06 mag.

\item[$r_d$] - scale length of the disk derived from a fit to the one-dimensional I-band photometric profile. 
\end{description}

\subsection{Photometric quantities from the 2MASS XSC} \label{xsc}

The photometric quantities in the XSC are derived by inspecting the 2MASS images 
with the extended source processor (GALWORKS), which was
developed for this purpose (see Jarrett \etal ~2000). By the time the 2MASS images get to GALWORKS a point source spread function has already been derived for each image. 
GALWORKS identifies extended sources (galaxies as well as other more nearby resolved sources) and measures a variety of parameters for each one as
described in \citet{J00} and in the 2MASS Explanatory Supplement \citep{cutri}. Very large galaxies (greater than 5\arcmin ~in size at $\sim 20$th \magarc ~in J-band) are first identified in in a separate large galaxy atlas\footnote{See: \url{http://irsa.ipac.caltech.edu/applications/2MASS/LGA/}} and then added to the XSC. The quantities from the XSC that we use are:

\begin{description}
\item[$(a/b)_{\rm J}$] - J-band axial ratio fit to the 3$\sigma$ isophote (which is close to the 20th {\rm mag\,arcsec$^{-2}$} isophote). The center of this ellipse fit is fixed to the peak intensity pixel and since only one isophote is fitted, it is an approximation to the ellipticity of the galaxy. If the 3$\sigma$ isophote is too close to the center of the galaxy the bulge might interfere with the derived ellipticity, making the galaxy appear more face-on than it really is. Of the three 2MASS bands, the J axial ratio should suffer the least from this effect; we note that no systematic offset is seen between the J-band and I-band axial ratios, as can be seen Figure \ref{avcol}d. An estimate of the average error of 0.1 in $\log(a/b)$ is derived from the scatter in this plot; this is more than three times as large as the average error estimated for the equivalent I-band quantity.

\item[$r_{21}^{\rm J}$, $r_{21}^{\rm H}$, $r_{21}^{\rm K}$] - semi-major axis of the isophotal elliptical aperture set to the 21st {\rm mag\,arcsec$^{-2}$} isophote individually in each band. An alternative to this would have been the semi-major axis of the ellipse fit to the 20th \magarc ~isophote, which is the preferred aperture in the XSC \citep{J03}. In Section 5.1 it is shown that \ri ~in J-band is at $\sim 2.5$ scale lengths of the disk ($r_d$), while in H and K\s ~it is $\sim 3.3 r_d$. A similar analysis for $r_{20}$ shows that $r_{20}^{\rm J} \sim 1.6 r_d$, and r$_{21}^{\rm H,K} \sim 2.2 r_d$. We choose to use \ri ~in order to trace the disks as far out as possible (which still is not far compared to $r_{23.5}^I \sim 3.5 r_d$). Since the typical 1 $\sigma$ background noise is 21.4, 20.6 and 20.0 mag in J, H and K\s ~respectively \citep{J03}, we must be careful in using the radii set to the 21st \magarc ~level, especially in H and K\s. In J-band \ri ~should be reasonably well defined, and does indeed show good correlation with $r_{20}$ with a scatter of about 3 \arcsec. The same analysis for H and K\s ~shows a similar correlation with a slightly larger scatter (3.5\nolinebreak \arcsec ~and 4 \nolinebreak\arcsec~ respectively). When $r$ is smaller than 7\nolinebreak\arcsec ~or greater than 120\nolinebreak\arcsec ~(for either isophote and in all bands) it does not follow the correlation well. These radii are not used.

\item[J$_{21{\rm f}}$, H$_{21{\rm f}}$, K$_{21{\rm f}}$] - magnitudes within the 21st \magarc ~elliptical isophote set in J-band (referred to as ``J fiducial'' are used to construct colors because of the necessity of having a consistent aperture. The uncertainty on these magnitudes listed in the XSC is that associated with Poisson noise and background subtraction and does not include other errors such as the zero point calibration and the uncertainty in the elliptical isophotal aperture (see Jarrett et al. 2000 for more details). This formal uncertainty is small (with the largest quoted at only 0.2 mag) and the real uncertainty is likely to be larger. The ``Level One Science Requirements'' of 2MASS (which can be found in the on-line Explanatory Supplement) require that the extended source photometry be accurate and uniform to within 10\%
, and we assume this is a closer estimate of the error in the magnitudes.

\item[J$_{21{\rm f}}$, H$_{21{\rm f}}$, K$_{21{\rm f}}$] - magnitudes measured within the 21st \magarc ~level set in J-band are used to construct colors.

\item[J$_{\rm ext}$, H$_{\rm ext}$, K$_{\rm ext}$] - XSC ``extrapolated'' magnitudes designed to represent the total magnitude of the galaxy. 
These magnitudes are derived from an extrapolation of the radial surface brightness profile from the 20th \magarc ~elliptical isophote down to $r_{\rm tot}$ which is set to 5 times the scale length from the J-band fit. See \citet{J03} for more details.

\end{description}

\subsection{Corrections}

 In order to use the photometry from the 2MASS XSC to its best potential for our purposes, various corrections for seeing, galactic extinction and cosmological redshift need to be applied. These corrections have already been applied to the I-band quantities in a similar manner as that described below (for details, 
see Haynes \etal ~1999a).
 
\subsubsection{Seeing correction} \label{corrections}

\begin{figure*}
\plotone{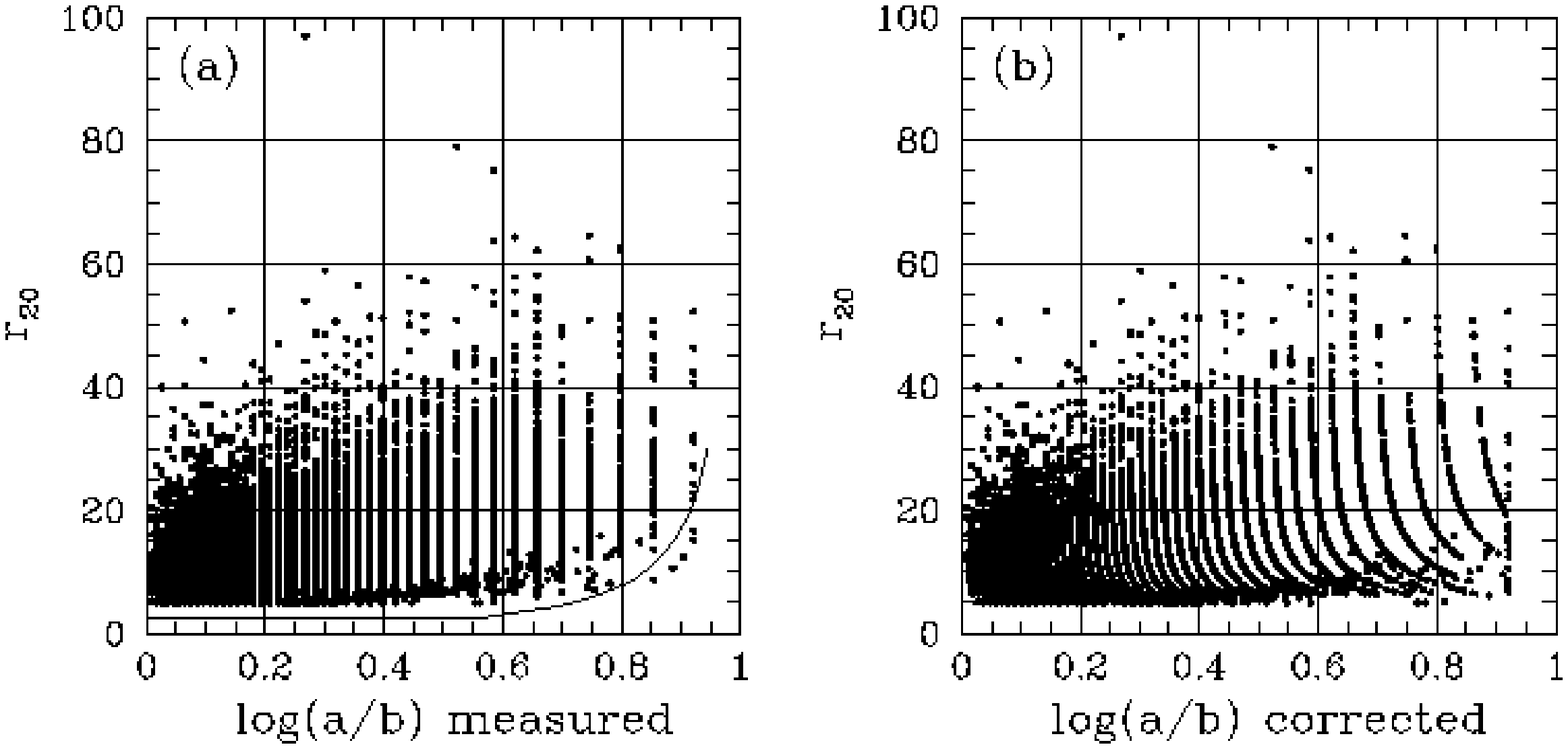}
\caption{Isophotal angular semi-major axis (in arcsec) at the 20th \magarc ~level plotted against the log of the axial ratio (which is measured at about the 20th \magarc ~level). (a) As observed; the line shows the theoretical limit from our simulations. (b) Corrected for seeing with FWHM= 2.5\arcsec ~as described in text. The obvious banding is due to the accuracy of tabulated values in the XSC.}
\label{seeing}
\end{figure*}

Figure \ref{seeing}a shows a plot of $\log(a/b)_{\rm J}$ vs. the 20th \magarc ~isophotal radius (the approximate isophotal radius at which $a/b$ is reported). We see that there is a lack of galaxies with both high axial ratio and small size. This is 
due mainly to the resolution of the images set by the seeing and the pixel size. 

In the XSC, seeing is characterized using a modified exponential of the form
\begin{equation}
    f(r) = f(0) \exp \left[-\left(\frac{r}{\alpha}\right)^{1/\beta}\right]
\end{equation}
fit to many stars in each frame, where $f(0)$ is the central surface brightness, $r$ is the radius in arcseconds and $\alpha$ and $\beta$ are free parameters. The ``radial shape'' (${\rm sh} = \alpha \times \beta$) is reported,  and an approximate relation is given in the 2MASS Seeing and Image Statistics \footnote{2MASS Seeing and Image Statistics (Cutrie 1998) is an on-line document found at: \url{http://spider.ipac.caltech.edu/staff/roc/2mass/}} to recover the full width at half maximum of the point spread function;
\begin{equation}
{\rm FWHM (arcseconds)} = 3.13~ {\rm sh} - 0.46.
\end{equation}
The mean value of this FWHM is about 2.5\arcsec ~for the galaxies in our samples (the 2MASS raw images have a pixel size of 2\arcsec ~which sets the lower limit on the resolution), and the spread is small. We derive a correction by numerically convolving exponential disks of varying inclination and angular sizes with a Gaussian of FWHM = 2.5\arcsec. Axial ratios are measured at a radial distance from the galaxy center of $1.7 r_d$ (where $r_d$ refers to the scale length of the exponential disk), corresponding to $r_{20}$ in J-band (the ratio $r_d/r_{20}$ is obtained in the same way as $r_d/r_{21}$ in Section 5.1). The derived correction is parameterized as
\begin{equation}
(a/b)_{\rm corr} = (a/b)_{\rm obs}(1 - 0.02x + 0.21x^2 - 0.01x^3)
\end{equation}
with 
\begin{equation}
x = {\rm FWHM} \left ( \frac{a/b}{r_{20}}\right )_{\rm measured}
\end{equation}
(where FWHM = 2.5\arcsec ~is used in all cases and $r_{20}$ is measured in 
arcseconds in the J-band image). Hereafter, when we refer to the J-band axial 
ratio, we refer to the quantity corrected by the above equation. The result of 
the correction can be seen in Figure \ref{seeing}b. In a few cases where the galaxy is very small and highly inclined the correction is large so to avoid over-correcting a maximum value of $a/b = 1/0.12$ is imposed and these galaxies are assumed to have an inclination of 90\arcdeg. From Figure 1b it appears that the correction removes much but perhaps not all of the bias, so very small galaxies may appear less inclined than they really are. It should also be noted that the derived correction assumes light profiles are exponential, which may not be an ideal approximation within the $r_{20} \sim 1.7 r_d$ radius in J-band where the contribution from a bulge may be significant.

\subsubsection{Galactic Extinction and K-corrections}

Magnitudes are corrected for galactic (\ie ~originating within the Milky Way) extinction using the DIRBE galactic extinction maps (Schlegel, Finkbeiner \& Davis 1998).  The column densities listed in the map are converted
to extinction in the relevant bands by $A_J = 0.902 E(B-V)$, $A_H = 0.576  E(B-V)$ and $A_K = 0.367 E(B-V)$. A small cosmological k-correction is also included using a simple linear fit to the low redshift end of the models of \citet{pog} which is $k_J = -0.68z$, $k_H = -0.40z$, and $k_K = -1.52z$.

The magnitude in band X corrected for galactic extinction and cosmological k-correction is then
\begin{equation}
m_{\rm X} = m^{\rm obs}_{\rm X} - A_{\rm X} + k_{\rm X}.
\end{equation}

The isophotal radii also need to be corrected for galactic extinction. If we assume the disk photometric profiles are exponential with scale length $r_d$ in band X, then
\begin{equation}
r^{\rm iso}_{\rm X} = r^{\rm obs}_{\rm X} - r_d \frac{A_{\rm X}}{1.086}
\end{equation}
(since the surface brightness at radius $r$ in an exponential disk is given by $\mu(r)=\mu(0)+1.086~r/r_d$). We also correct sizes for a cosmological stretch factor $(1+z)^2$.

\section{Reddening}\label{colour}

\begin{figure*}
\plotone{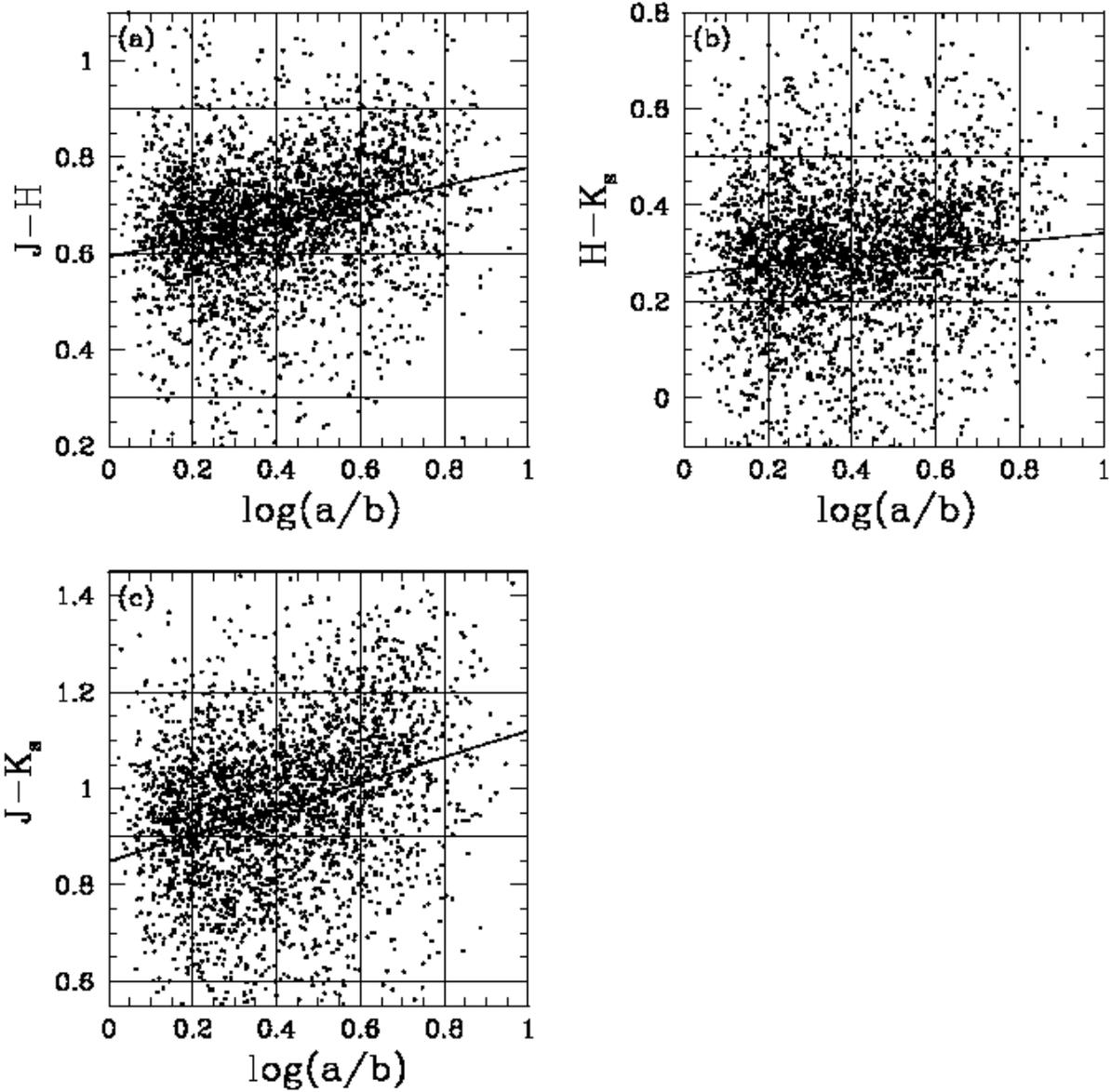}
\caption{NIR color plotted against the $\log$ of the axial ratio (measured from I-band images) for 3035 galaxies for which I-band photometry is available (SFI2 sample). The slopes of the linear fits are 0.2 in J-H (a), 0.1 in H-K$_{\rm s}$ (b), and 0.3 in J-K$_{\rm s}$ (c). The formal error in determining the slope is 0.01 mag in all cases, while the scatter in the relationship is roughly 0.1 mag. Assuming there is no extinction in K$_{\rm s}$-band (which will be tested below) this gives a lower limit to the value of $\gamma$ (where $\Delta M = \gamma \log(a/b)$) in J and H bands of 0.3 and 0.1 mag respectively. Note the indication of non-linearity in this relationship (especially in J-K$_{\rm s}$ (c)) which is much more evident in Figure \ref{avcol} and is discussed in Sections 4, 6.3 and 7.}
\label{colour-inc}
\end{figure*}

In this section we use the samples defined above in Section 2 to investigate the effects of internal extinction in the disks by investigating color gradients with inclination.
This test has the advantage of suffering little from selection effects.
For the SFI2 sample, we use the I-band axial ratios as the measure of inclination since these have errors almost three times smaller than the equivalent J-band measure from the XSC. The result is shown in Figure \ref{colour-inc} in which colors (measured within an elliptical aperture with a semi-major axis of \ri ~in J-band) are plotted vs. axial ratio. It is clear that reddening is present in all colors, which sets a lower limit to the extinction in the J and H bands by assuming that the extinction in K$_{\rm s}$ is negligible. Following G94 and others, we parameterize the internal extinction as
\begin{equation}
\Delta M = \gamma \log (a/b).
\end{equation}
The lower limit on $\gamma$ is then 0.3 mag in J and 0.1 mag in H. The formal error in fitting this slope to the data is 0.01 mag in both cases, while the scatter in the relationship is 0.1 mag. If there is extinction present in K$_{\rm s}$-band (\ie ~$\gamma_{\rm K} \neq 0$), the actual values of $\gamma$ in J and H will be $0.3 + \gamma_{\rm K}$ and $0.1 + \gamma_{\rm K}$ respectively. An indication that the linear law of Eqn 7 may not be appropriate for the highest inclinations is apparent in panels (a) and (c) of Figure 2, an issue we'll revisit below and in Section 6.

\begin{figure*}
\plotone{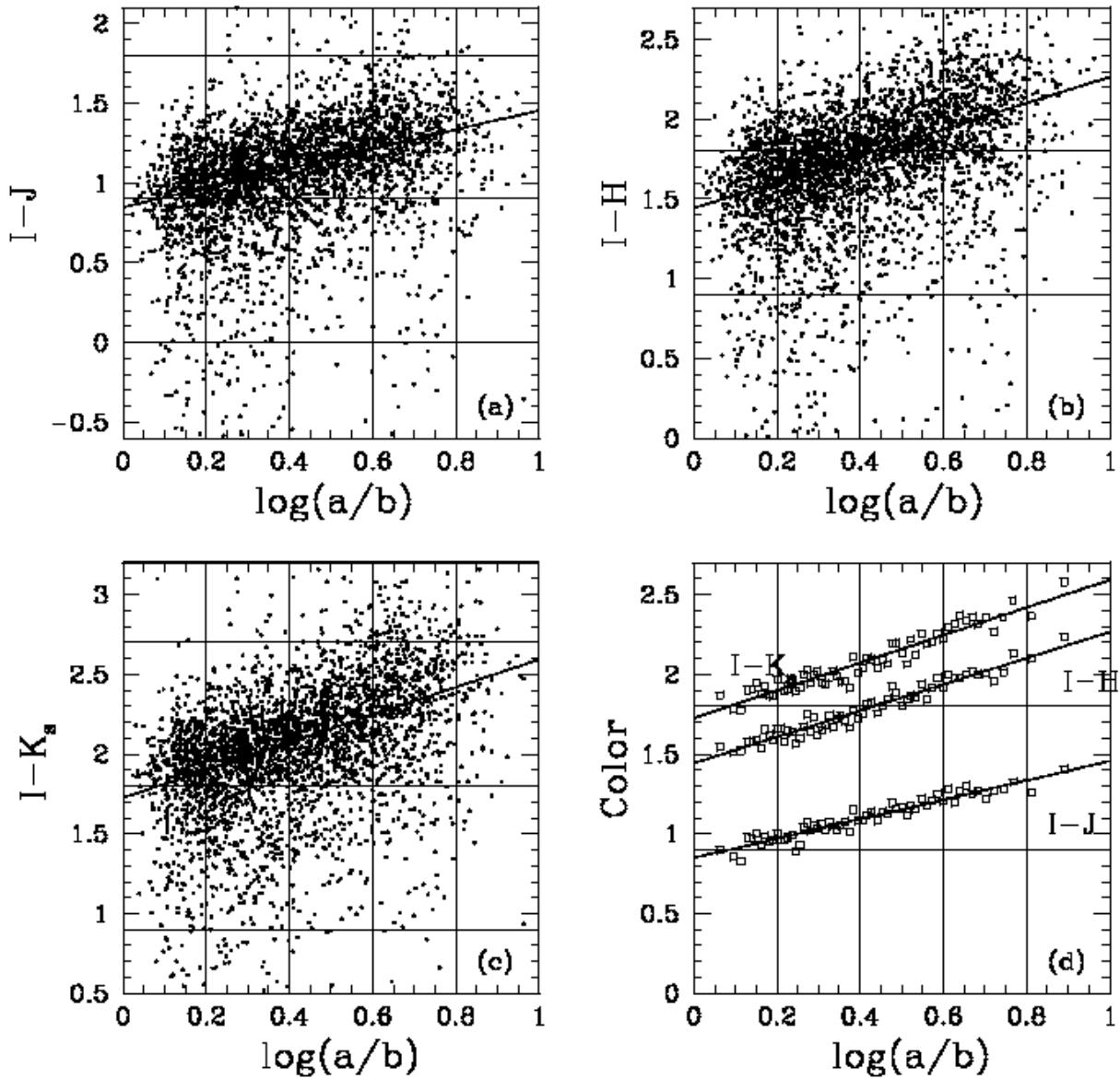}
\caption{NIR color plotted against the $\log$ of the axial ratio (measured from the I-band images) for 3035 galaxies for which I-band photometry is available (SFI2 sample). Note that the aperture in which the I-band magnitude is measured is not the same as the J, H and K$_{\rm s}$  apertures, but this should not affect the slope only the offset. The lines shown are fits to the median in bins of $\log(a/b)$ which is also shown in plot (d). The slope, $\gamma$ is 0.73 in I-J (a), 0.82 in I-H (b) and 0.94 in I-K$_{\rm s}$ (c). The scatter is 0.4 mag in all cases, and the error on determining the slope from the medians is 0.03 mag. Assuming no extinction in K$_{\rm s}$-band this then gives a lower limit to the extinction in I of 0.9 magnitudes, consistent with the value of $\gamma = 1.05 \pm 0.08$ derived in G94 and $\gamma \simeq 1.0$ in \citet{t98}. For comparison with Fig \ref{colour-inc} the vertical scale is exactly three times larger.}
\label{icol}
\end{figure*}

\begin{figure*}
\plotone{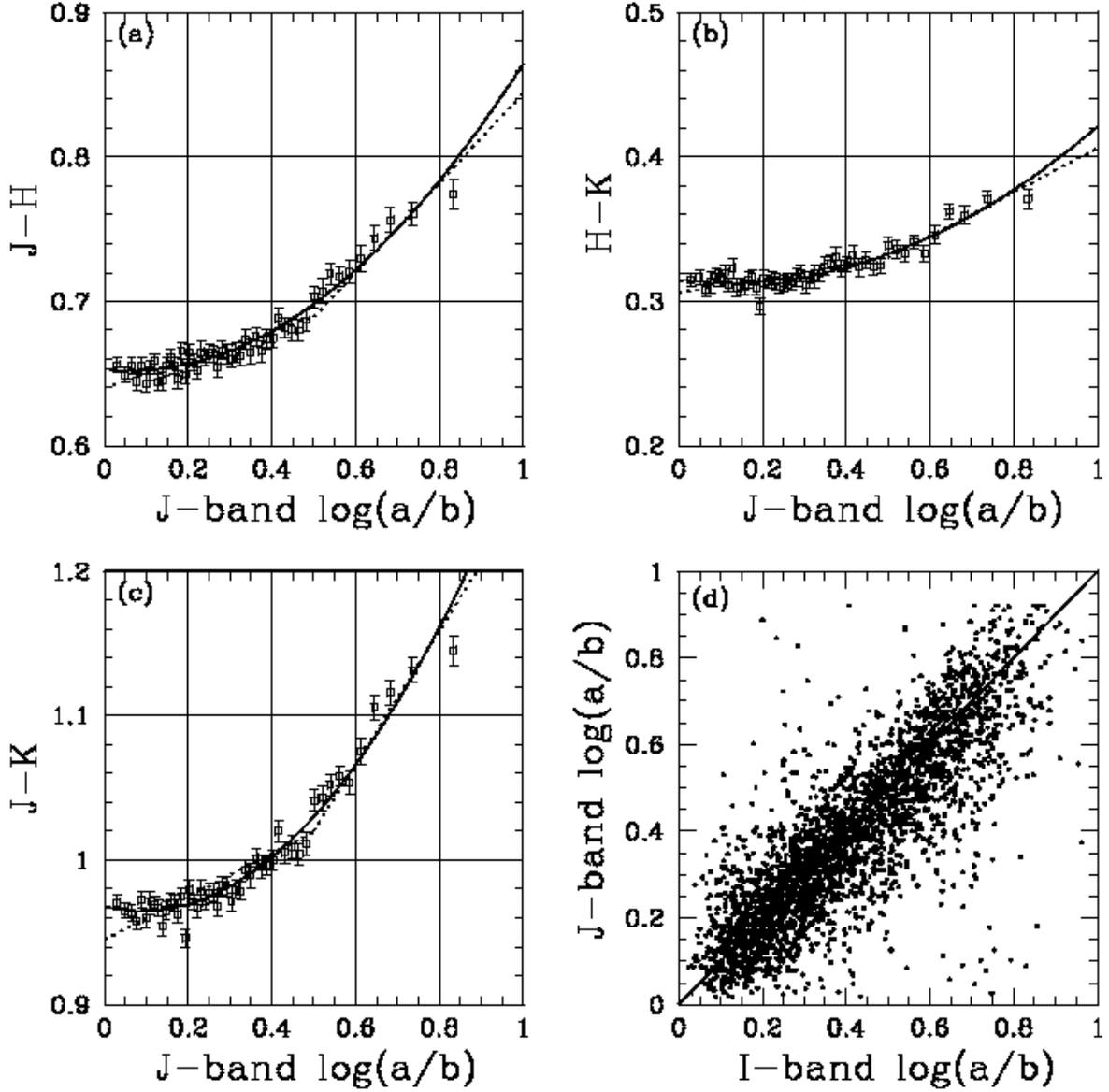}
\caption{NIR color plotted against the $\log$ of the axial ratio (measured from the J-band images) for 15224 galaxies (AGC$z$ sample) binned in groups of 300 (error bars show the statistical error on determining the mean). 
The fit to the points is given as a quadratic (solid line) and as a bi-linear relation with a break at $\log(a/b) = 0.5$ (dotted line). 
Panel (d) shows $\log(a/b)$ in J vs. $\log(a/b)$ in I (both corrected for seeing) for the 3035 SF2 galaxies in the sample; the scatter is 0.1 in $\log(a/b)$ and there is no significant indication of bias in the J-band axial ratios. For comparison with Figure \ref{colour-inc} the range in ordinate value in (a), (b) and (c) is exactly three times smaller. }
\label{avcol}
\end{figure*}

For the SFI2 sample, I-band magnitudes are also available. These magnitudes can be 
used to construct I-J, I-H and I-K$_{\rm s}$ colors, with the understanding that the apertures are not consistent between I and the other bands. This mismatch should 
not, in the first order, affect the slope of the reddening, just the absolute color. The results of this are shown in Figure \ref{icol}. Under the 
assumption that the extinction in K$_{\rm s}$  is negligible, we derive a lower limit on the extinction in I-band of $\gamma = 0.94 \pm 0.03$ mag (where the error is the formal error in determining the slope). G94 find an extinction law of $\gamma_I = 1.05 \pm 0.08$, while \citet{t98} derive $\gamma_I \simeq 1.0$, suggesting that if the extinction in K$_{\rm s}$ is not 
equal to zero, it is likely to be small.

For the much larger AGC$z$ sample of galaxies (which does not have I-band photometry for all the galaxies), we can plot the J-H, J-K$_{\rm s}$ and H-K$_{\rm s}$ colors against the J-band axial ratios (Figure \ref{avcol}). Since the sample is large, we combine the points into averages in $\log(a/b)$ bins which makes the inadequacy of a simple linear law for the relationship (as noted in Figure \ref{colour-inc}) stand out. 
We fit two separate lines respectively for $\log(a/b)$ less than and greater than 0.5, and also include a quadratic fit for comparison.
The parameterization of $\Delta M$ with a linear law in Eqn 7 is an expedient form, and internal extinction can well depart from such a linear law, depending on the details of the relative distribution of stars and dust (see, e.g. Evans 1992 and G94). In most simulations, in fact, $\Delta M$ steepens with increasing $\log(a/b)$ as will be shown for a simple photometric model in Section \ref{model}.  
Our adoption of bi-linear and quadratic fits reflects a least departure from simple expediency. The location of the break in the bi-linear law at $\log(a/b) =0.5$ for all color trends is also practical but arbitrary. The slopes of the bi-linear relation are 0.10 $\pm$ 0.02 and 0.31 $\pm$ 0.05 for J-H, 0.05 $\pm$ 0.02 and 0.15 $\pm$ 0.05 for H-K$_{\rm s}$, and 0.15 $\pm$ 0.02 and 0.46 $\pm$ 0.05 for J-K$_{\rm s}$. The quadratic fit is 0.65(20) - 0.03(1)$\log(a/b)$ + 0.24(2)$[\log(a/b)]^2$ for J-H, 0.31(25) - 0.04(2)$\log(a/b)$ + 0.14(2)$[\log(a/b)]^2$ for H-K$_{\rm s}$, and 0.97(28) - 0.07(2)$\log(a/b)$ + 0.39(3)$[\log(a/b)]^2$ for J-K$_{\rm s}$.

\section{Effect of inclination on isophotal radius} \label{rads}

\subsection{Formal corrections for inclination} 
 
 If a galaxy is completely opaque within a given isophotal radius in the face-on perspective then no change in the measured radius is expected 
with inclination. If it is assumed that the face-on disk is transparent at that radius, then the isophotal radius should increase as the disk is tilted and we look through a longer path length. The expected change derived for the simple case of a pure exponential disk is
\begin{equation}
r_{iso} = r_{iso}^{\circ} + 2.302 r_d^{\circ} \log(a/b)
\end{equation}
where $r_{iso}^{\circ}$ and $r_d^{\circ}$ are respectively the isophotal radius and the disk scale length in the face-on case. If we 
assume that the scale length changes with inclination like $r_{iso}$ (which is 
justified by the observation that the ratio $r_d/r_{21}$ does not change with inclination, as seen below) then we derive
\begin{equation} \label{delta}
\frac{r_{iso}}{r_{iso}^{\circ}} = 1 + 2.302 \frac{r_d}{r_{iso}} \log(a/b).
\end{equation}

We have the scale lengths in I-band measured for the SFI2 sample and make the assumption that scale lengths do not change significantly between I, J, H and K$_{\rm s}$-band, as found by \citet{deJongII} from a sample of 86 face-on spiral galaxies. In their sample, scale length is seen to decrease as the wavelength increases, but not significantly between I and K. A measure of the scale length is provided in the XSC for all of the galaxies in our sample, but given the notorious difficulties with fitting scale lengths, we choose not to use that automated measure here, and only use scale lengths from the SFI2 sample fit ``by hand'' (see \citet{h99a} for more details on the SFI2 scale lengths).
Using these scale lengths we estimate $r_d/r_{21} \nolinebreak = \nolinebreak 0.4\nolinebreak\pm\nolinebreak0.1,~0.3\pm0.1$ and $0.3\pm0.1$ in J, H and K$_{\rm s}$ for the SFI2 galaxies (where the error here is the scatter in the sample), and we assume that these ratios hold for all the galaxies in the AGC$z$ sample. The ratio $r_d/r_{21}$ in K$_{\rm s}$  may be slightly smaller than in H, but this is within the errors. No significant dependence of the ratio on inclination
is observed in any band. We note that the $r_d/r_{21}$ ratios imply that $r_{21}^J$ is at only $2.5 \pm 0.6$ scale lengths out in the disk, while in H and K$_{\rm s}$, $r_{21}$ is still at only $3.3\pm 1.1$ scale lengths. The commonly reported isophotal radius at I-band is $r_{23.5}$ at 3.5-4 scale lengths which is further out than $r_{21}^{\rm J}$. 

If the face-on disks were transparent at $r_{\rm 21}$ we can rewrite Eqn \ref{delta} as
\begin{equation}
r_{21}/r_{21}^{\circ} = 1 + \delta \log(a/b) 
\end{equation}
where $\delta = 0.9 \pm 0.3, 0.7 \pm 0.2$ and $0.7 \pm 0.2$ in J, H and K$_{\rm s}$ respectively, when we use the  $r_d/r_{21}$ values given above. The isophotal radius does not change ($\delta = 0$) if the disk is opaque at $r_{21}$ so $\delta$ must decrease as the disk 
becomes less transparent. The main limitation on Eqn \ref{delta} is that it assumes that the photometric profile is exponential, while it is expected that the importance of the bulge light increases in these NIR bands.

\subsection{Are disks transparent at $r_{21}$?} \label{tausection}

\begin{figure*}
\plotone{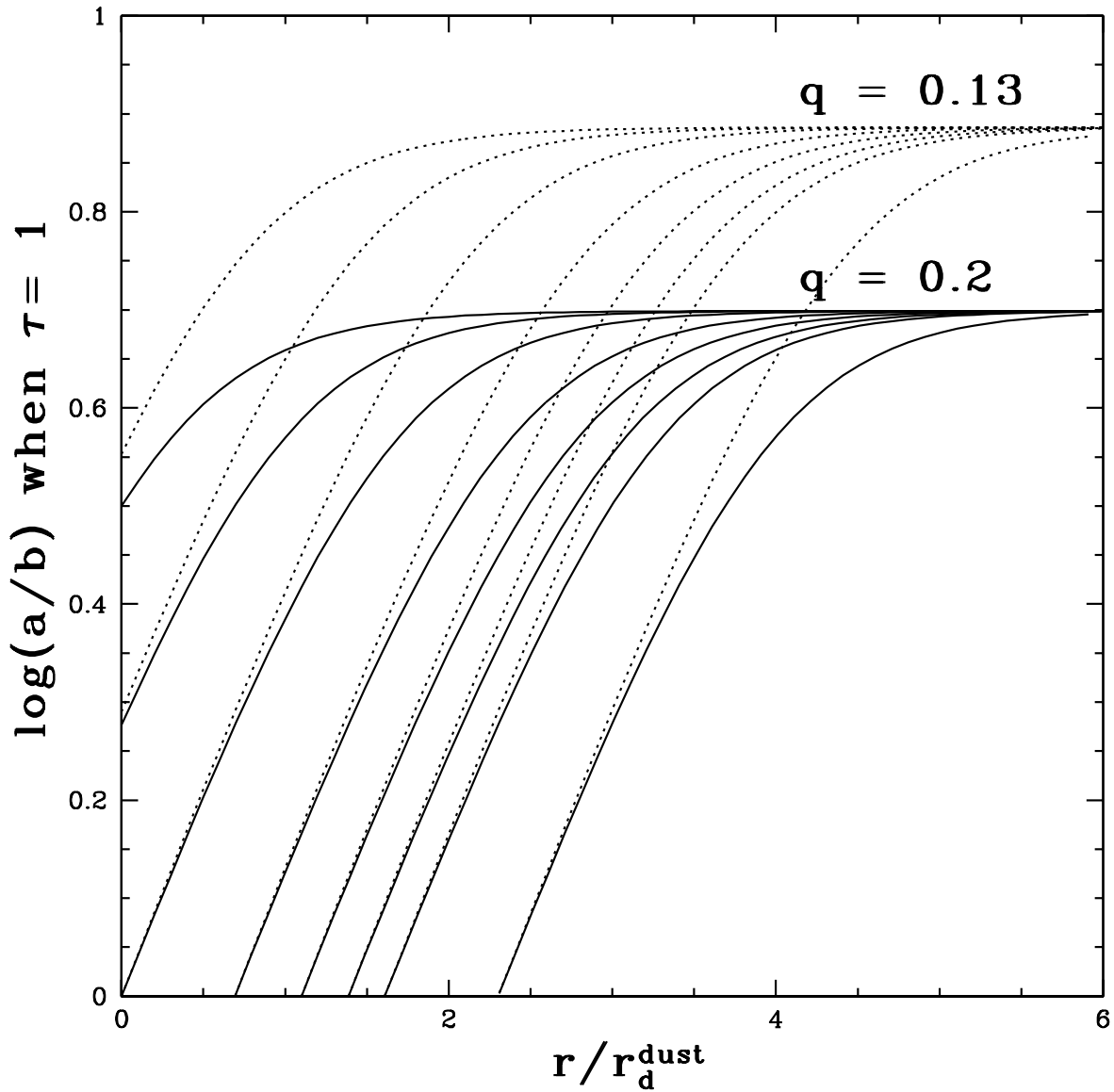}
\caption{The observed axial ratio (assuming an intrinsic axial ratio, $q= 0.13$ for the dotted lines or 0.2 for the solid lines, where $\cos^2 i = ((b/a)^2 - q^2)/(1-q^2)$) at which a galaxy with a pure exponential disk of dust (with $r_d^{\rm dust} = r_d^\star$) becomes opaque is shown as a function of galactic radius for various central optical depths (0.25, 0.5, 1, 2, 3, 4, 5, 10 from left to right). 
Far from the center,
the disk is transparent at most inclinations, but further in the disk can become optically thick for fairly modest $\tau^\circ(0)$ at high enough inclinations.}
\label{tau}
\end{figure*}

Here we test the often made assumption that the outer disks of galaxies 
are transparent, especially in the NIR. Suppose that stars and dust are exponentially distributed with scale lengths $r_d^{\rm dust} = r_d^\star = r_d$. The disk opacity at radius $r$ and inclination $i$ is then
\begin{equation}
\tau(r,i) = \frac{\tau^{\circ}(0) \exp (-r/r_d)}{\cos i}
\end{equation}
where $\tau^{\circ}(0)$ is the opacity at $r=0$ in the face-on perspective. There have been many attempts to derive $\tau^{\circ}(0)$ for spiral galaxies. G94 derive an upper limit in I-band of $\tau_I^{\circ}(0) < 5$ using statistical arguments with a sample of about 2000 Sc galaxies. A lower estimate of $\tau^{\circ}(0)$ was obtained by \citet{xi99}: they claim that $\tau^{\circ}(0) < 1$ in 7 edge-on spiral galaxies they model in K, J, I, V and B-bands. \citet{kuchinski98} derive values of $\tau^{\circ}(0)$ using BVRIJHK photometry and radiative transfer models for 15 spiral galaxies. They adopt values in the ranges of 0.15-0.6, 0.1-0.4, and 0.05-0.2 for J, H and K respectively. \citet{moriondo} find $\tau^{\circ}(0)=$0.3-0.5 in H-band from a sample of 154 galaxies. Even if $\tau_I^{\circ}(0)$ were as high as $5$ it is easy to see that at $r_{23.5,{\rm I}} \sim 3.5 r_d^\star$ the disk is close to transparent out to quite high inclinations. However it has been shown above that 
$r_{21}$ in J, H and K$_{\rm s}$ is closer to the center of the disk than $r_{23.5, {\rm I}}$ so (in the above model) even for relatively low values of $\tau^{\circ}(0)$ the opacity could become  significant in a highly inclined disk. Figure \ref{tau} shows the inclination at which the galaxy becomes optically thick ($\tau = 1$) plotted as a function of the radial distance for $\tau^{\circ}(0)$ varying between 0.25 and 10.

At $r_{21}$ in J-band, which is $2.5 \pm 0.6$ $r_d$, the disk can become opaque before it is fully inclined for even the smaller $\tau^{\circ}(0)$. 
If we interpret the change of slope for high inclinations shown in Figures \ref{avcol} and \ref{holmJH}a as the onset of $\tau \geq 1$ at $r_{21}$, a ``break'' for J-band ($r_{21, {\rm J}} \sim 2.5 r_d$) and H and K\s ~($r_{21,{\rm H/K}} \sim 3.3 r_d$) near $\log(a/b) \sim 0.5$ would 
suggest that the disk opacity in the J, H and K\s ~bands, while low, is non-negligible and that even at \ri ~an important fraction of the light is extincted when disks are edge-on.

Recent studies suggest that the dust scale length is larger than that of the stars, see for example \citet{moriondo} who use $r_d^{\rm dust} = \xi r_d^\star$ with $\xi = 1.5$. In the above model this simply changes the position of \ri with respect to the dust scale length so that for example $r_{21, {\rm J}} \sim 2.5 r_d^\star \sim (2.5/\xi) r_d^{\rm dust}$. In that case a lower central face-on opacity is required to explain the behavior.

\subsection{A statistical test for the inclination dependence of isophotal radius}

It is possible to test the mean change in isophotal radius with inclination in a statistical way using large samples of galaxies. For the test described below, the biases introduced by the selection function of the sample can be quite severe. In order to keep track of these effects, we use only galaxies north of declination $= -3$\arcdeg ~from the AGC$z$, since that region is known to be more homogeneously and completely sampled in the catalog.
Most of the galaxies in the AGC were originally identified by eye from photographic plates on which small, dim galaxies are easier to pick out if they are edge-on. In an attempt to remove this bias we also restrict the sample to B $<$ 15.5. The samples are then further reduced to be flux limited (using $m_{\rm ext}$ from the XSC) or angular size limited (using \ri) within each band. 
The final 6 subsamples are taken to represent complete flux or angular size limited samples, and are summarized in Table \ref{subsamples}.
\begin{deluxetable}{ccc}
\tabletypesize{\small}
\tablecaption{Samples for the study of $r_{\rm iso}$ and $m_{\rm ext}$ vs. $\log(a/b)$. 
\label{subsamples}}
\tablewidth{0pt}
\tablehead{
\colhead{Band} &  \colhead{Limit} & \colhead{No. galaxies}}
\startdata
J & $m_{\rm ext} < 11.5$ mag & 1774 \\
J & $r_{21} > 20$ \arcsec & 3120 \\
H & $m_{\rm ext} < 11$ mag& 2274 \\
H & $r_{21} > 25$ \arcsec & 3137\\
K$_{\rm s}$ & $m_{\rm ext} < 10.5$ mag & 1733\\
K$_{\rm s}$ & $r_{21} > 30$ \arcsec & 2377\\
 \enddata
\end{deluxetable}

\begin{figure*}
\plotone{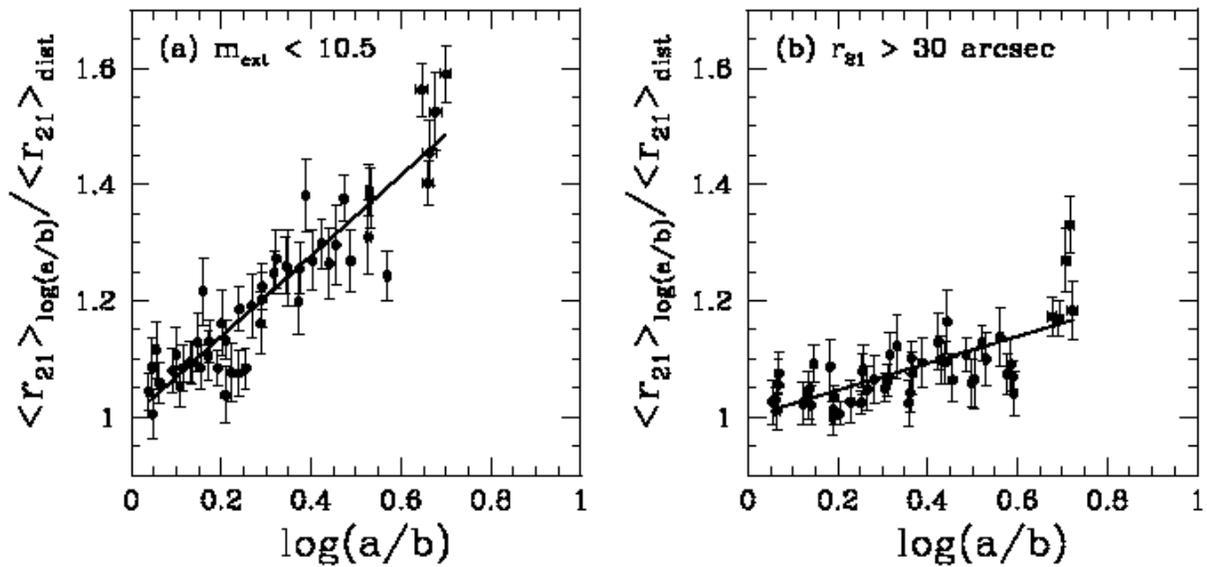}
\caption{Averaged values of the isophotal radius, $r_{21}$ in K$_{\rm s}$-band in bins of $\log(a/b)$ scaled by the value of all the galaxies in the same distance bin (as described in Section 5.3). Fit to this is a function $c_1 + c_2\log(a/b)$, the plot shows the data and the fit divided by $c_1$. The slope, $\delta = c_2/c_1$ is 0.7 $\pm$ 0.1 in the magnitude limited sample (1733 galaxies) and 0.2 $\pm$ 0.1 in the radius limited sample (2377 galaxies), where the error is the formal error on determining the slope.}
\label{avtest_r}
\end{figure*}

For either size or flux-limited samples, the mean value of $r_{\rm iso} = r_{21}$, expressed in kpc, will increase with distance. In order to remove this simple manifestation of the Malmquist bias, we divide our sample
into five distance groups with the same number of galaxies in each.  Within each group the mean of $r_{21}$ is then computed in bins of $\log (a/b)$. These bins all have an equal number of galaxies in them ($\sim$30-60 depending on the size of the subgroup). Since there are few galaxies at high inclinations this 
restricts analysis of subgroups to $\log(a/b) <$ 0.7-0.8 in most cases. In each of the bins the values of $<r_{21}>_{\log(a/b)}$ are scaled by the mean of the value for the whole distance group ($<r_{21}>_{\rm dist}$). A relation $c_1 + c_2\log(a/b)$ is fit to the points and the value of $\delta$ (see Eqn 10) is calculated as $c_2/c_1$. The results in K$_{\rm s}$-band  for both the size and flux-limited sub-samples are shown in Figure \ref{avtest_r}, and are summarized for each band in Table \ref{radiusaverage} along with the values of $\delta$ derived in Section 5.1 for the case of an exponential disk which is transparent at $r_{21}$.

A comparison of panels 6a and 6b, constructed from flux and angular size-limited samples respectively, well illustrates the amplitude of the bias introduced by the character of the sample. This is explained as follows; as disks become more inclined, on average $r_{21}$ moves out and the total magnitude dims (assuming that the galaxies are neither totally transparent throughout the whole disk or totally opaque at \ri). In a sample defined by a threshold angular size, $r_{\rm min}$, galaxies that in the face-on perspective would be smaller than $r_{\rm min}$ (and thus excluded from the sample) at high inclinations will appear to have a size greater than $r_{\rm min}$ and will be included in the sample. As a result, this sample will exhibit a shallow $r_{\rm iso}$ v. $\log(a/b)$ slope as shown by Figure 6b.
Alternatively, in a magnitude limited sample, intrinsically larger galaxies fall below the threshold magnitude if they are edge-on due to the dimming caused by extinction. This will have the effect of steepening the slope, as seen in Figure 6a. 

\begin{deluxetable}{cccc}
\tabletypesize{\small}
\tablecaption{Measured and estimated values of $\delta$.
\label{radiusaverage}}
\tablewidth{0pt}
\tablehead{
\colhead{Band} &  \colhead{Flux limited} & \colhead{Angular size limited} & \colhead{Transparent exponential disk}}
\startdata
J & 1.2$ \pm$ 0.1 & 0.5 $\pm$ 0.1 & 0.9 $\pm$ 0.3\\
H & 0.8$ \pm$ 0.1 & 0.3 $\pm$ 0.1 & 0.7 $\pm$ 0.2 \\
K$_{\rm s}$ & 0.7$ \pm$ 0.1 & 0.2 $\pm$ 0.1 & 0.7 $\pm$ 0.2 \\
 \enddata
\end{deluxetable}

Note that if disks were opaque, \ri ~would not change with inclination ($\delta=0$), and this result would be recovered by the analysis of an angular size-limited sample. On the other hand, if the disk dimming due to internal extinction is small or negligible, a flux-limited sample will return the correct value of $\delta$ through an analysis such as that represented in Figure 6a. The presumption of very low dimming at K\s-band, supported by the results discussed in Section 4, is further supported by the similarity of the value of $\delta$ in columns 1 and 3 of Table 3 for K\s-band. A final choice on the value of $\delta$ in each band needs to be made after an estimate of the amount of internal extinction so will be discussed in Section 6.3. It is probably safe to assume there is very little extinction in K\s ~so we can conclude here that $\delta_K = 0.7 \pm 0.1$. 

\section{Effect of inclination on magnitude}

\subsection{Formal corrections for inclination}

 Note that if the disks of galaxies are completely transparent all the way to the center, then there should be no change in the total magnitude with inclination (\ie ~$\gamma = 0$) , while the magnitude enclosed within a fixed isophote will brighten as the physical radius at which it is measured moves out. If the disks are completely opaque, then both the total and isophotal magnitudes will dim by an amount 2.5 $\log(a/b)$. 

\subsection{A statistical test for the inclination dependence of total magnitude}

We test for the inclination dependence of the total magnitude (here taken to be $m_{\rm ext}$ as described in Section 3.3) using a method similar to that described in Section 5.3. We fit a relation $\Delta M = \gamma \log(a/b)$ to the mean values of the magnitude binned in $\log(a/b)$ within five separate distance groups. We use the flux and angular size limited samples described in Section 5.3, since (as for the test in Section 5.3) the interpretation must take into account bias introduced by the selection function of the sample. For an angular size limited sample with threshold $r_{\rm min}$, if the disk is not completely opaque all the way through, then $\gamma$ will be increased as intrinsically small and dim galaxies move above $r_{\rm min}$ at high inclinations. If the disks are completely opaque, this type of sample would recover the exact result of $\gamma=2.5$ as the angular sizes are unaffected by inclination. For a flux limited sample, and if the galaxies are not completely transparent, internal extinction will dim galaxies at higher inclinations and force $\gamma$ to be smaller. If the disks are completely transparent, this sample would recover the exact result of no change of total magnitude with inclination. The change in $\gamma$ with sample selection is observed in all photometric bands explored here (as summarized in Table \ref{avmag}), providing further evidence that galaxies are neither completely opaque nor completely transparent in these bands. We note that under the assumption of little extinction in K\s, the flux-limited sample in that band should give a value of $\gamma$ very close to the real value. This implies $\gamma_K = 0.2 \pm 0.1$. In the other bands, J and H, we use this test to provide upper and lower limits to the actual extinction law as parameterized by $\gamma$. 

\begin{deluxetable}{ccc}
\tabletypesize{\small}
\tablecaption{Measured values of $\gamma$. 
\label{avmag}}
\tablewidth{0pt}
\tablehead{
\colhead{Band} &  \colhead{Flux limited} & \colhead{Angular size limited}}
\startdata
J & 0.2$ \pm$ 0.1 & 1.2 $\pm$ 0.1 \\
H & 0.1$ \pm$ 0.1 & 0.8 $\pm$ 0.1 \\
K$_{\rm s}$ & 0.2$ \pm$ 0.1 & 0.7 $\pm$ 0.1 \\
\enddata
\end{deluxetable}

\subsection{The Modified Holmberg Test} \label{holmberg}

 Following G94 we use a modified version of the Holmberg test (Holmberg 1958, 1975).
Define $\Sigma^\prime = m_{\rm iso} + 5 \log r_{\rm iso}$ as the surface brightness within an elliptical aperture with semi-major axis $r_{\rm iso}$. The original Holmberg test consists of measuring the change of $\Sigma^\prime$ with disk inclination.
Unfortunately a similar trend can be found in both opaque and transparent galaxies. If the galaxy is totally opaque we expect that the radius at which a given isophote is measured will not shift with inclination and that the magnitude will dim by 2.5 $\log(a/b)$, so a slope of 2.5 is expected. In the fully transparent case, fixed brightness isophotes will move to larger radii as the galaxies are tilted which will also decrease $\Sigma^\prime$. From Eqn. 8 this geometric dilution alone (\ie ~not taking into account the change in the isophotal magnitude as the aperture changes) is given by
\begin{equation}
5 \log(r_{21,a}/r_{21,b}) = 5 \log  \left [ \frac{1 + 2.302(r_d^{\circ}/r_{21}^{\circ}) \log(a/b)_a}{1 + 2.302(r_d^{\circ}/r_{21}^{\circ}) \log(a/b)_b}\right ]
\end{equation}
between axial ratios $(a/b)_a$ and $(a/b)_b$. This would give a change in $\Sigma^\prime$ between $\log(a/b)$ = 0 and 1 of 1.1 $\pm$ 0.3 for the  $r_d/r_{21}$ derived in Section 5.1 in H and K$_{\rm s}$ and 1.4 $\pm$ 0.3 in J (where the errors indicate the scatter in $r_d/r_{21}$). These values provide an upper limit to the change since the isophotal magnitude will brighten slightly as the aperture increases. 

Here, we do not use the isophotal magnitude, but rather a measure of the total magnitude (extrapolated magnitude from the XSC), to remove the effects of the physical size of the aperture changing with inclination. $r_{\rm iso}$ is respectively $r_{21,{\rm J}}$, $r_{21,{\rm H}}$ and $r_{21,{\rm K}}$ for the three 2MASS bands. In the following discussion we will refer to this radius as \ri, with the understanding that it is separately measured in each of the three bands. The slope fit to $\Sigma^\prime = m_{\rm ext} + 5 \log r_{21}$ vs. $\log(a/b)$ for galaxies in the AGC$z$ sample with lower inclinations (below $\log(a/b) = 0.5$) is 2.0 $\pm$ 0.1 in J, 1.6 $\pm$ 0.1 in H and 1.3 $\pm$ 0.1 in K$_{\rm s}$. Since this is higher than the values given above for the transparent case we take it to be indication that there is extinction present in the inner parts of these galaxies, while the outer parts are largely transparent. This is supported by the radial variation of colors and surface brightness seen in large galaxies in the XSC \citep{J03}. If the disk is transparent at the edges but not in the center, the change of $\Sigma^\prime = m_{\rm ext} + 5 \log r_{\rm iso}$ from face-on to edge-on will be larger than the values for geometric dilution alone and can in fact be larger than 2.5 as there will be some dimming of the magnitude which is dominated by light from the inner parts of the disk.

\begin{figure*}
\plotone{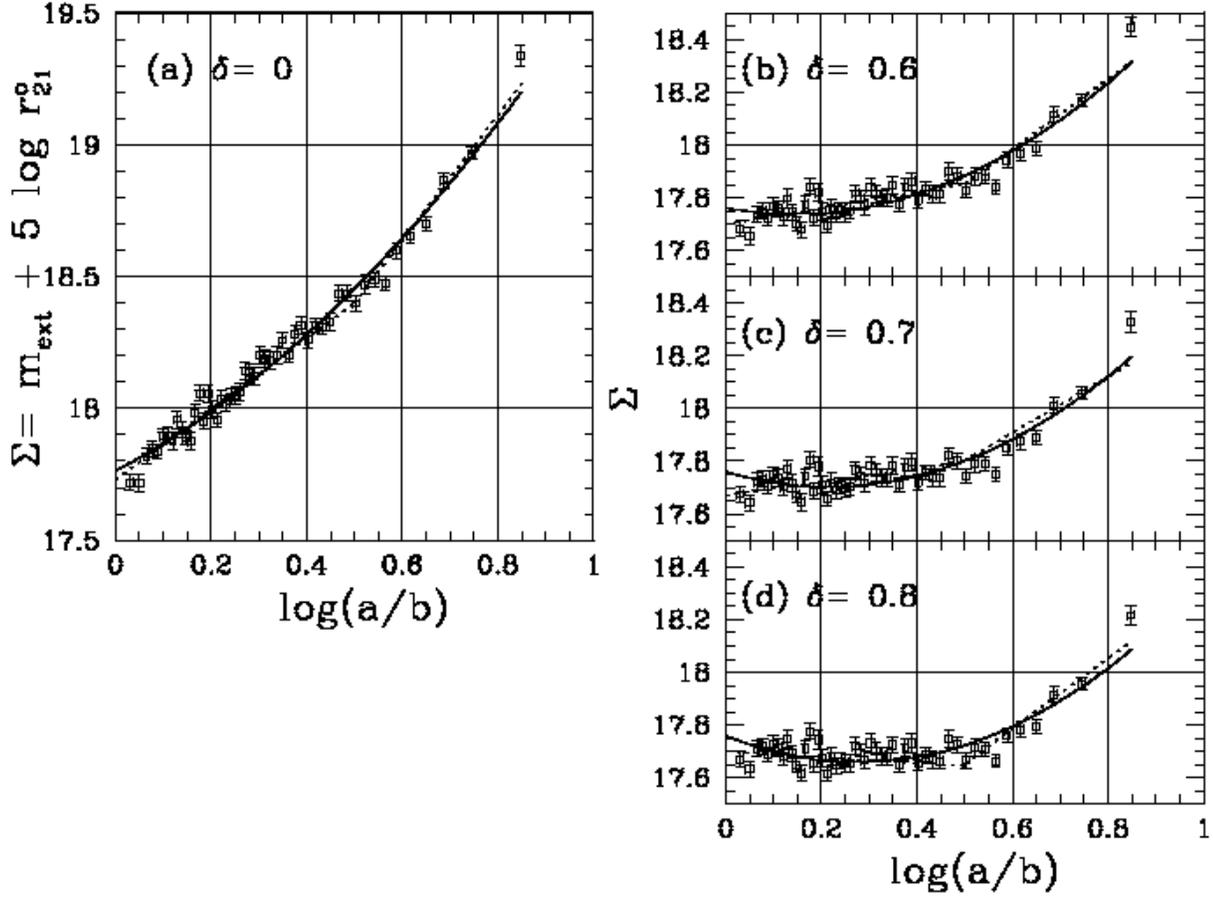}
\caption{Holmberg test in K\s-band (see Section 6.3). The values of $\delta$ shown are that used to correct the isophotal radius to the face-one value using $r_{21}/r_{21}^{\circ} = 1 + \delta \log(a/b)$. 
A bi-linear fit is shown (dotted line) with the break at $\log(a/b) = 0.5$. The fitted slopes are (a) 1.34 and 2.38, (b) 0.31 and 1.34, (c) 0.26 and 1.06, and (d) 0.05 and 1.07. The error on the slope is on the order of 0.1 for the low inclinations and 0.2 for the high inclinations. Also shown is a quadratic fit for comparison (solid line). This smooth function may be more desirable, but the unphysical upturn at low $\log(a/b)$ may make the bi-linear law more useful.
Points are averages in bins of $\log(a/b)$ with roughly 300 galaxies per bin, where the error bars show the error in determining the mean in the bin.}
\label{holmK}
\end{figure*}

In order to remove the effect introduced by an inclination dependence of the isophotal radius, we adopt a 
modified version of the Holmberg test in which the isophotal radius in the face-on view is derived via Eqn 10. The value of $\delta$ which is used in Eqn 10 can significantly change the resulting slope. This is illustrated in Figure \ref{holmK} which shows the standard Holmberg test (for galaxies observed in K$_{\rm s}$-band) as well as the test with corrections using $\delta$ = 0.6, 0.7 and 0.8 (spanning the error range for $\delta_{\rm K}$ discussed in Section 5). 

\begin{figure*}
\plotone{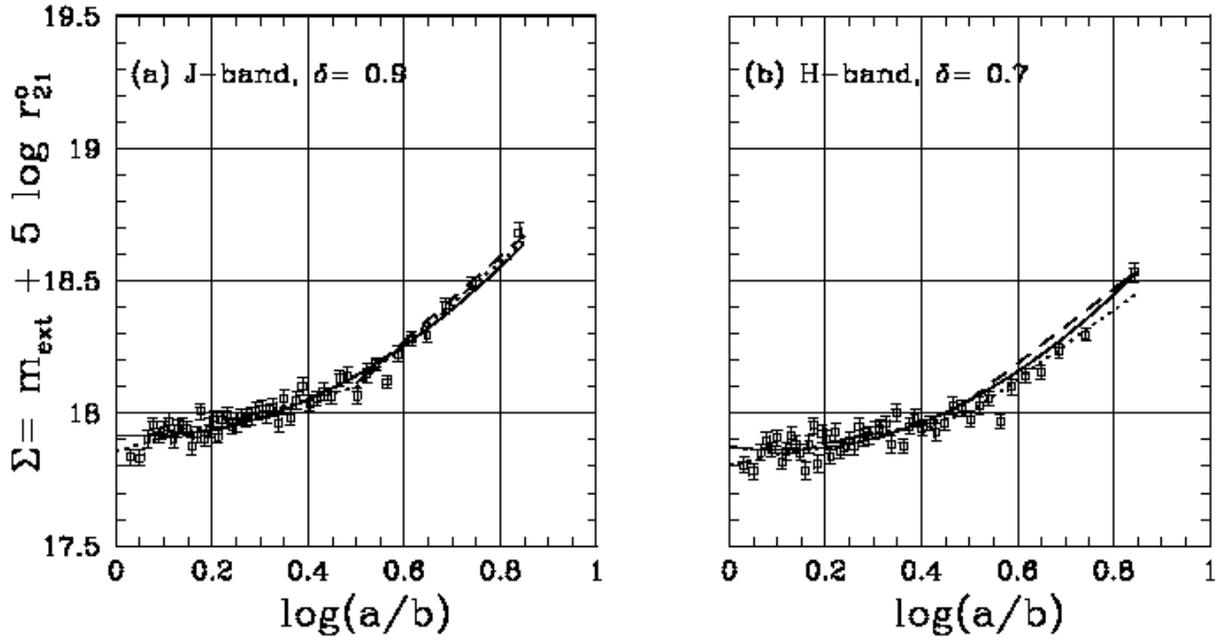}
\caption{Modified Holmberg test in J and H-bands with the chosen value of $\delta$ (see Section 6.3). Shown is a bi-linear fit to both cases where the break is at $\log(a/b) =0.5$ (dotted line). The slopes are 0.48 $\pm$ 0.1 and 1.57 $\pm$ 0.2 in J-band (a), and 0.39 $\pm $ 0.1 and 1.30 $\pm$ 0.2 in H-band (b). Also shown (long dashed line) is the adopted correction. The solid lines show a quadratic fit for comparison.}
\label{holmJH}
\end{figure*}

We will discuss the obvious up turn of these relationships further in Section \ref{model} but for now deal with the low inclination ($\log(a/b) <  0.5$) and high inclination ($\log(a/b) \geq 0.5$) regions separately.
At the low inclinations in K$_{\rm s}$-band,
$\gamma$ changes from 0.31 (with $\delta = 0.6$) to 0.05 ($\delta=0.8$); a significant difference within the plausible range of $\delta$. As a consequence of this interdependency, the best values of $\delta$ and $\gamma$ must be decided upon together. Considering all the available evidence (Sections 5.1, 5.3 and 6.2) we adopt $\delta = 0.7$ in K$_{\rm s}$-band, $\delta = 0.7$ in H-band, and $\delta = 0.9$ in J-band. As a comparison, \citet{moriondo} derive $\delta_{\rm H} = 0.85 \pm 0.16$ at the same isophotal level (from their sample of 154 galaxies). The results for just these values of $\delta$ in J and H can be seen in Figure \ref{holmJH}, while the test with the chosen value of $\delta$ in  K$_{\rm s}$-band is shown in Figure \ref{holmK}c. We thus derive $\gamma_{\rm J} = 0.48 \pm 0.15$, $\gamma_{\rm H} = 0.39 \pm 0.15$, and $\gamma_{\rm K} = 0.26 \pm 0.15$ for $\log(a/b) < 0.5$ from this test (where the error indicates the variation in the value within the adopted value of $\delta \pm 0.1$.
Similar values of $\gamma_H = 0.16\pm 0.22$ \citep{moriondo} and $\gamma_{\rm K} \simeq 0.22$ \citep{t98} have been found before, although \citet{graham01} and others suggest $\gamma_K \simeq 0$. 
While the variation of $\gamma$ with the wavelength band depends on the detailed form of the geometric mixing of stars and dust, it is interesting to note that, in the simple geometric case of a foreground absorbing screen model, the extinction curve of our galaxy would yield $\gamma_{\rm J} = 0.5$,  $\gamma_{\rm H} = 0.3$, and $\gamma_{\rm K} = 0.2$, if $\gamma_{\rm I} = 1.0$ (see G94, Tully \etal ~1998, Schlegel \etal ~1998).

At the higher inclinations (above $\log(a/b)=0.5$) the modified Holmberg test is not appropriate as the disk is probably not completely transparent at \ri. The values of $\delta$ used to correct to the face-on value of \ri ~are therefore too large which artificially lowers the change in the average surface brightness and leads to an underestimate for $\gamma$. The standard Holmberg test (\ie ~$\delta=0$) does not work either. As discussed above, if there is significant extinction present in the inner parts of the disks where most of the light comes from, but the outer edges are still at least partly transparent, the slope of $\Sigma$ vs. $\log(a/b)$ can be larger than 2.5 and so is obviously an overestimate of $\gamma$. This is observed in J, H and K\s ~(the slopes are 2.9$\pm$0.2, 2.4$\pm$0.2 and 2.4$\pm$0.2 respectively), providing evidence that the most inclined galaxies are not totally opaque or transparent at \ri, and that there is significant extinction in the central regions. Since the extinction is lowest in K\s ~we expect that the disk is closest to transparent at \ri ~in this band and therefore adopt the slope of the modified Holmberg test here to give a tentative value of $\gamma_{\rm K}  \simeq 1.1 \pm 0.2$ for the more inclined galaxies (up to $\log(a/b) \sim 0.85$). We then use the slopes of the reddening in J-K\s ~and H-K\s ~(above $\log(a/b) = 0.5$: $\gamma_{\rm J} - \gamma_{\rm K} = 0.46\pm 0.05$ and $\gamma_{\rm H} - \gamma_{\rm K} = 0.31\pm 0.05$ from Figure 4.) to suggest $\gamma_{\rm J} \simeq 1.6\pm 0.2$ and $\gamma_{\rm H} \simeq 1.4\pm 0.2$. The difference between the bi-linear and quadratic fit is quite small (as can be seen in Figure \ref{holmJH}), so we also provide a smooth fit to the points giving $\gamma_{\rm J} = -0.12(15) + 1.14(15)\log(a/b)$, $\gamma_{\rm H} = -0.25(15) + 1.12(15)\log(a/b)$ and $\gamma_{\rm K} = -0.53(15) + 1.23(15)\log(a/b)$. Below $\log(a/b) = 0.1,0.2$, and 0.4 in J, H and K\s ~respectively the value for the correction should be zero -- the upturn in the quadratic fit at low inclination is obviously unphysical.

\section{Photometric Model} \label{model}
Here we investigate a simple photometric model which can be used to explain the upturn in the $\Delta M$ vs. $\log(a/b)$ relation at high $\log(a/b)$ seen in all three bands. The model used is that described in Section 9 of G94 which is based on the ``triple exponential" model of \citet{disney}. This model assumes that the dust is distributed in the disk with an exponential scale length equal to the stars and that the $z$ distributions are also exponential with a ratio $\zeta$ between the scale height of the dust and the stars. The face-on central surface brightness, $\tau^\circ(0)$ can be specified, as can the ratio, $q$, between the scale height and scale length of the stars. 
We use the ``standard galaxy" of G94 which has 7\% 
of the light in a central bulge.

\begin{figure}
\plotone{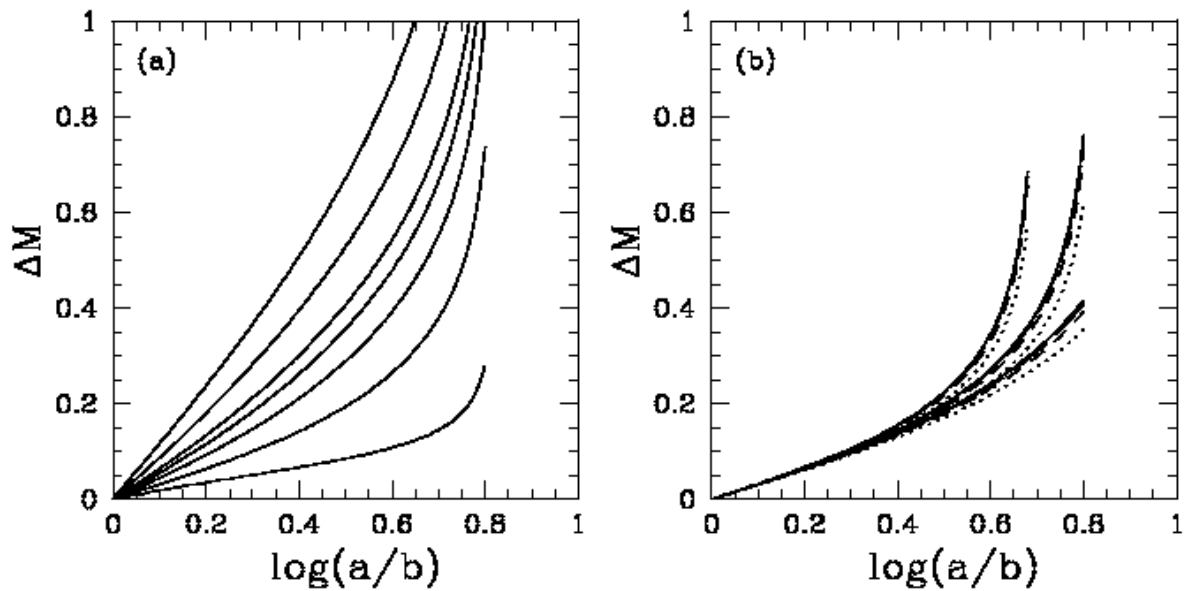}
\caption{Difference between the ``observed" magnitude and the face-on value from the ``triple exponential" model of \citet{disney} which assumes dust and stars are exponentially distributed with the same scale length. We vary the intrinsic axial ratio $q$, the ratio $\zeta$ between the scale heights of the dust and stars and the central face-on opacity, $\tau^\circ(0)$. (a) Results for $\zeta=0.5$, $q=0.15$ and $\tau^\circ(0)=0.1,0.5,1.0,1.5,2,4,10$ (from right to left). (b) Results for $\tau^\circ(0)=0.5$ and $q=0.1,0.15,0.2$ (right to left). The small effect of varying $\zeta$ between 0.1 (dotted line), 0.4 (short dashed), 0.7 (long dashed) and 1.0 (solid line) is also shown.}
\label{model1}
\end{figure}

Figure \ref{model1} shows the difference between the magnitude at a given axial ratio and the face-on magnitude calculated for various inputs to the model. In the model the upturn becomes more noticeable with increasing $q$ (\ie ~as the disk becomes thicker) and as the central face-on opacity decreases. The ratio between the thickness of the disk in the dust and stars has a much smaller effect as shown in Figure \ref{model1}b.

\begin{figure}
\plotone{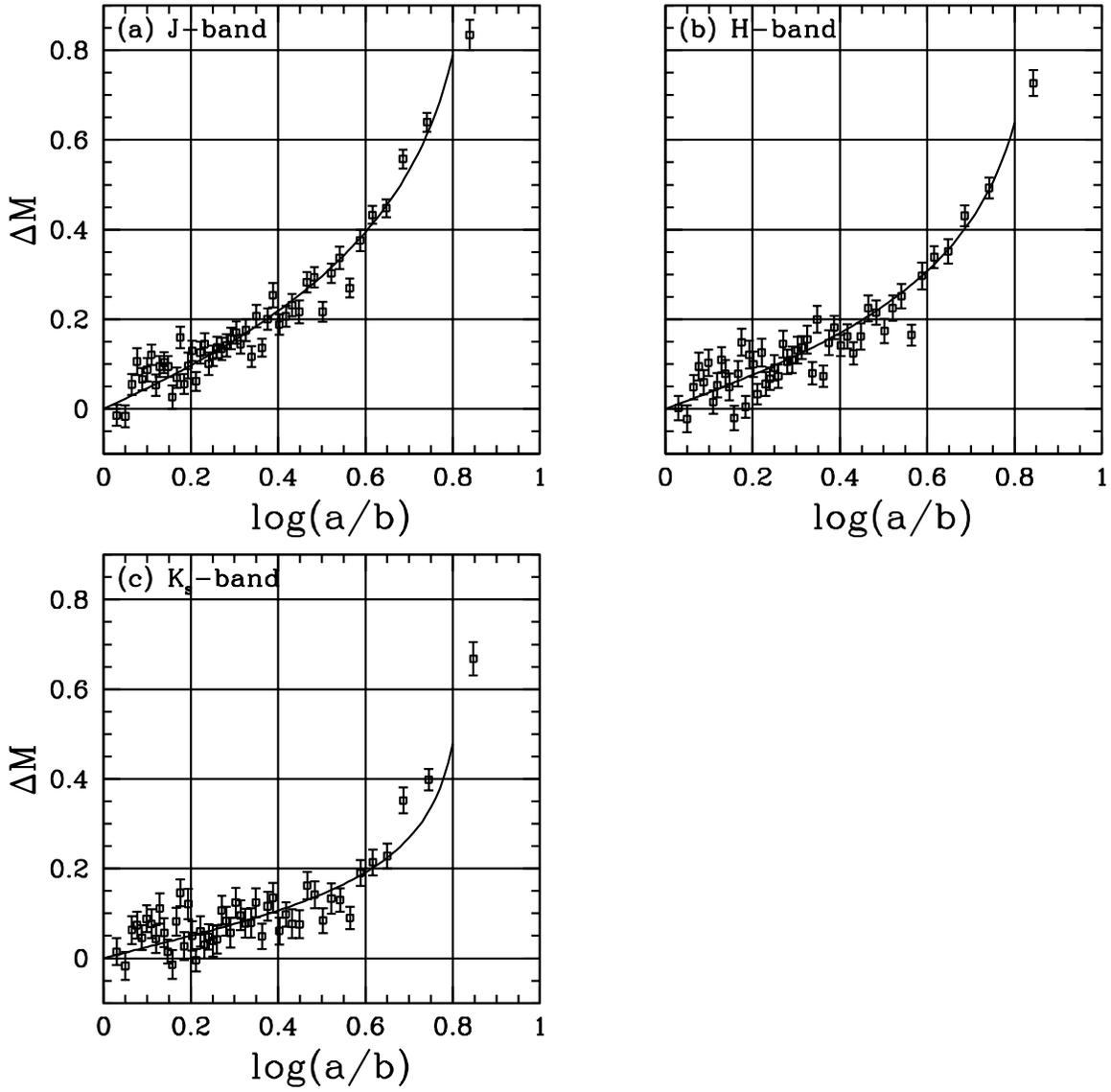}
\caption{Shown here are the data points from the Holmberg test in J-band (from Figure 8a) in (a), H-band (from Fig 8b) in (b) and K\s-band (from Fig 7c) in (c), offset to zero at $\log(a/b)= 0$. The solid lines show the photometric model which best fits (by eye) the data. All models are for $\zeta=0.5$. For the J-band data (a), $\tau^\circ(0) = 1.1$ and $q=0.14$; for H-band (b), $\tau^\circ(0) = 0.7$ and $q=0.14$ and for K\s-band (c), $\tau^\circ(0) = 0.3$ and $q=0.15$.}
\label{model2}
\end{figure}

Figure \ref{model2} shows our ``best fit" to the Holmberg test points (Section 6.3). Note that at high inclinations the points are artificially lowered (see discussion in Section 6.3). The simple photometric model described here can explain the observed behavior. While the dependence of the models on the dust-to-stars thickness ratio is weak, best fit models favor low values of $\zeta$. This is consistent with the expectation that stellar disks may appear somewhat thicker in NIR light, as they are dominated by an older stellar population. The models shown in Figure \ref{model2} were computed for $\zeta=0.5$. The other parameters of the model are $\tau^\circ(0) = 1.1$, $q=0.14$ in J-band, $\tau^\circ(0) = 0.7$, $q=0.14$ in H-band and $\tau^\circ(0) = 0.3$, $q=0.15$ in K-band. Reassuringly, the central face-on opacity of the best fit model decreases with increasing wavelength. Given the uncertainties in the Holmberg test (see Section 6.3) this model should be taken with caution, but does provide a plausible explanation for what is observed. 

\section{Luminosity dependence}

\begin{figure*}
\plotone{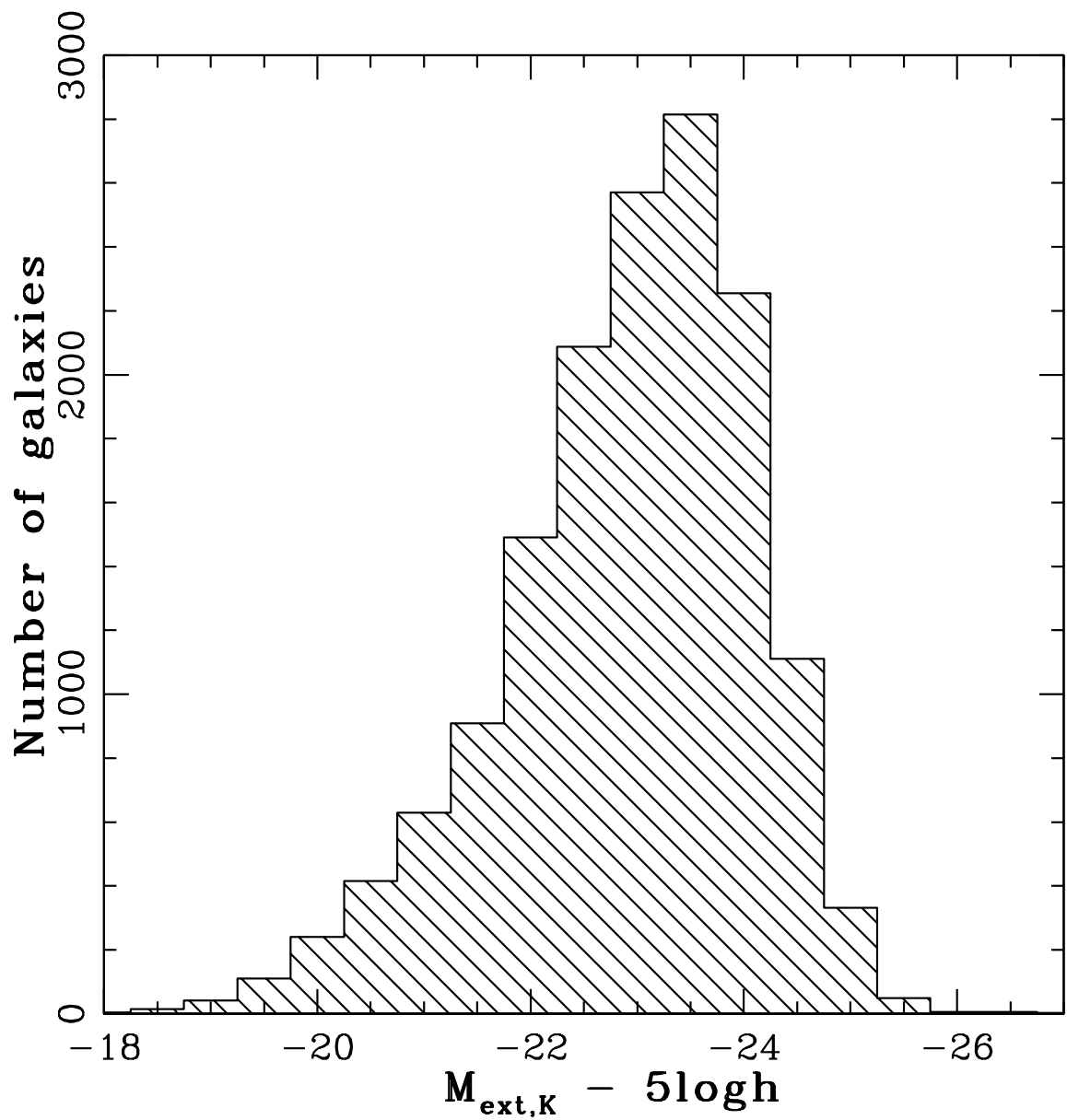}
\caption{Distribution of total magnitude in K\s-band of galaxies in the AGC$z$ sample.}
\label{lum}
\end{figure*}

Here we use the AGC$z$ sample of 15,244 galaxies to search for luminosity dependences of the extinction laws, as was done in both \citet{g95} and \citet{t98}. As a measure of the total luminosity of the galaxy we use $\MK$ in K-band since this is least affected by internal extinction. The luminosity distribution is shown in Figure \ref{lum}.  The peak of the distribution is at $\MK=-23$, which is near to $M_\star$ in K-band \citep{kochanek01} and most of the galaxies in the sample are within one magnitude of this.

The expected dependence is that brighter galaxies will show more internal extinction at a given inclination than dimmer ones. This is due both to the increase in physical size (and thus optical path lengths) and to the increase in metallicity (and thus presumably dust content) in the more luminous galaxies.

\subsection{Isophotal radius}
\begin{figure*}
\plotone{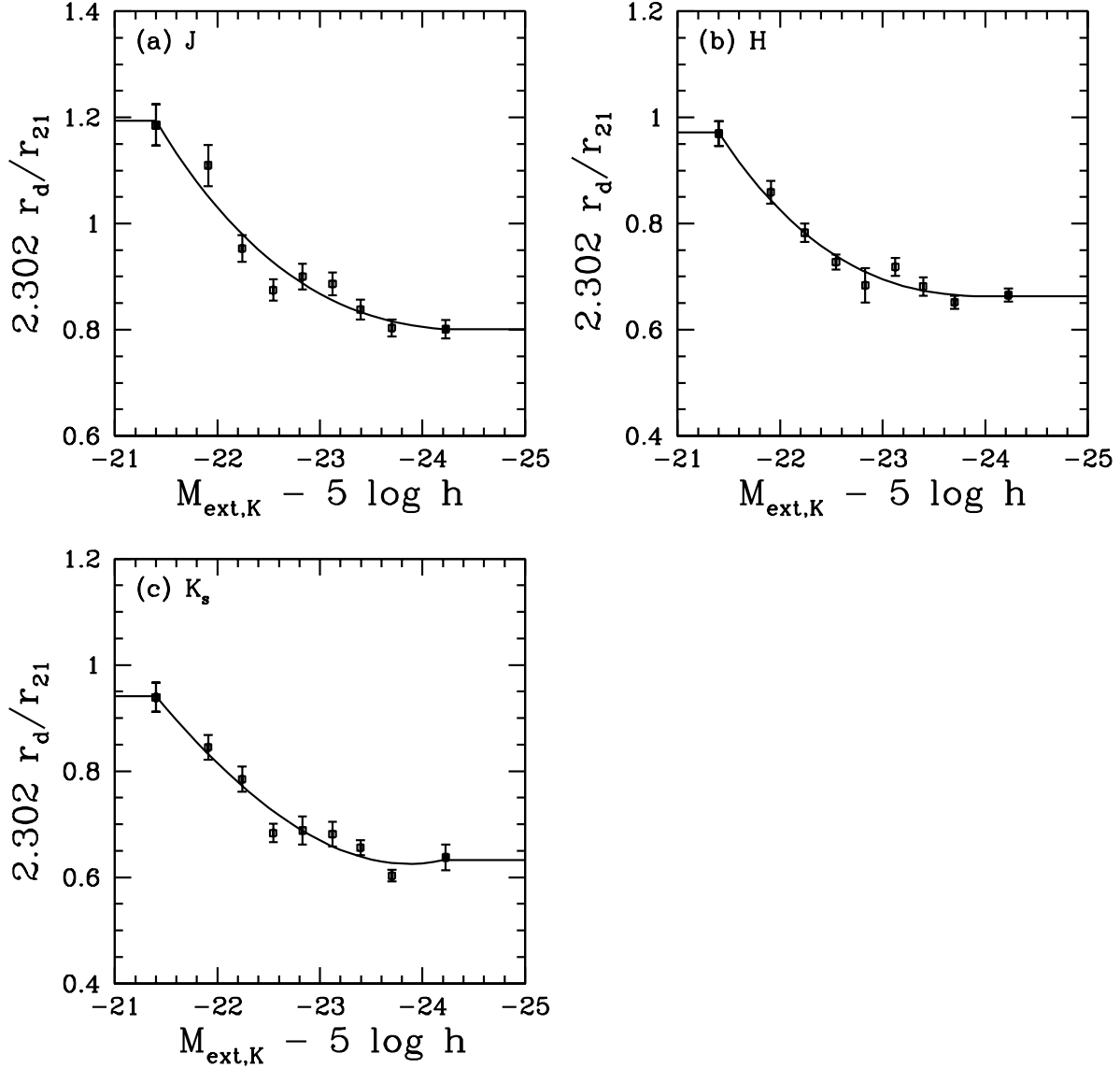}
\caption{The change of $2.302 r_d/r_{21}$ with luminosity for galaxies in the SFI2 sample. This parameter approximates $\delta$ if disk are nearly transparent at \ri. Data points show the averages in 9 bins of luminosity (with equal numbers of galaxies), while the solid lines show a cubic fit to the points truncated for the highest and lowest luminosity galaxies to be the value of the fit at $\MK = -21.4$ and -24.2 respectively. The fits are 
respectively for $\MK > -21.4$, between -21.4 and -24.2 and brighter than -24: $\delta = 1.19$, $\delta = 0.87 + 0.11x + 0.05x^2 + 0.007x^3$, $\delta =0.80 $ for J-band (a); $\delta=0.97$, $\delta = 0.70 + 0.07x + 0.05x^2 + 0.009x^3$, $\delta=0.66$ for H-band (b); and $\delta=0.94$, $\delta = 0.67 + 0.10x + 0.05x^2 - 0.004x^3$, $\delta=0.63$ for K\s-band (c). $x = \MK + 23$ in all cases.}
\label{del-lum}
\end{figure*}

Here we derive a parametric relation for the change of $\delta$ in $r_{21}/r_{21}^{\circ} = 1 + \delta \log(a/b)$ with total luminosity as measured by $\MK$. We discussed the expected value of $\delta$ for an exponential disk with scale length $r_d$ which is transparent at \ri ~in Section 5.1. This is given by $2.302 r_d/r_{21}$ and is shown by the points in Figure \ref{del-lum} for galaxies from the SFI2 sample in 9 bins of luminosity (each with an equal number of galaxies). There is a clear trend for the ratio $r_d/r_{21}$ (and thus $\delta$ if the disks are transparent at \ri) to decrease with luminosity, as was also found by \citet{g95} in I-band, who also discuss why this trend is to be expected. We fit a cubic to these points which we truncate at both the highest and lowest luminosities. These fits are shown by the solid lines in Figure \ref{del-lum} and the fit parameters are listed in Table \ref{results}. 

\subsection{Magnitude}
Here we apply the ``modified Holmberg test'' discussed in Section 6.3 to the full AGC$z$ sample to investigate the dependence of internal extinction (as parameterized by $\gamma$ in Eqn 7) on galaxy luminosity. We divide the sample into 10 luminosity bins with $\sim$ 1500 galaxies per bin. We correct the isophotal radius to the face-one view using 
the cubic fit to the points in Figure \ref{del-lum}.  We then fit a single linear law to $\Sigma$ vs. $\log(a/b)$ for each luminosity bin as the subdivision in luminosity bins restricts statistics, especially at high inclinations.

\begin{figure*}
\plotone{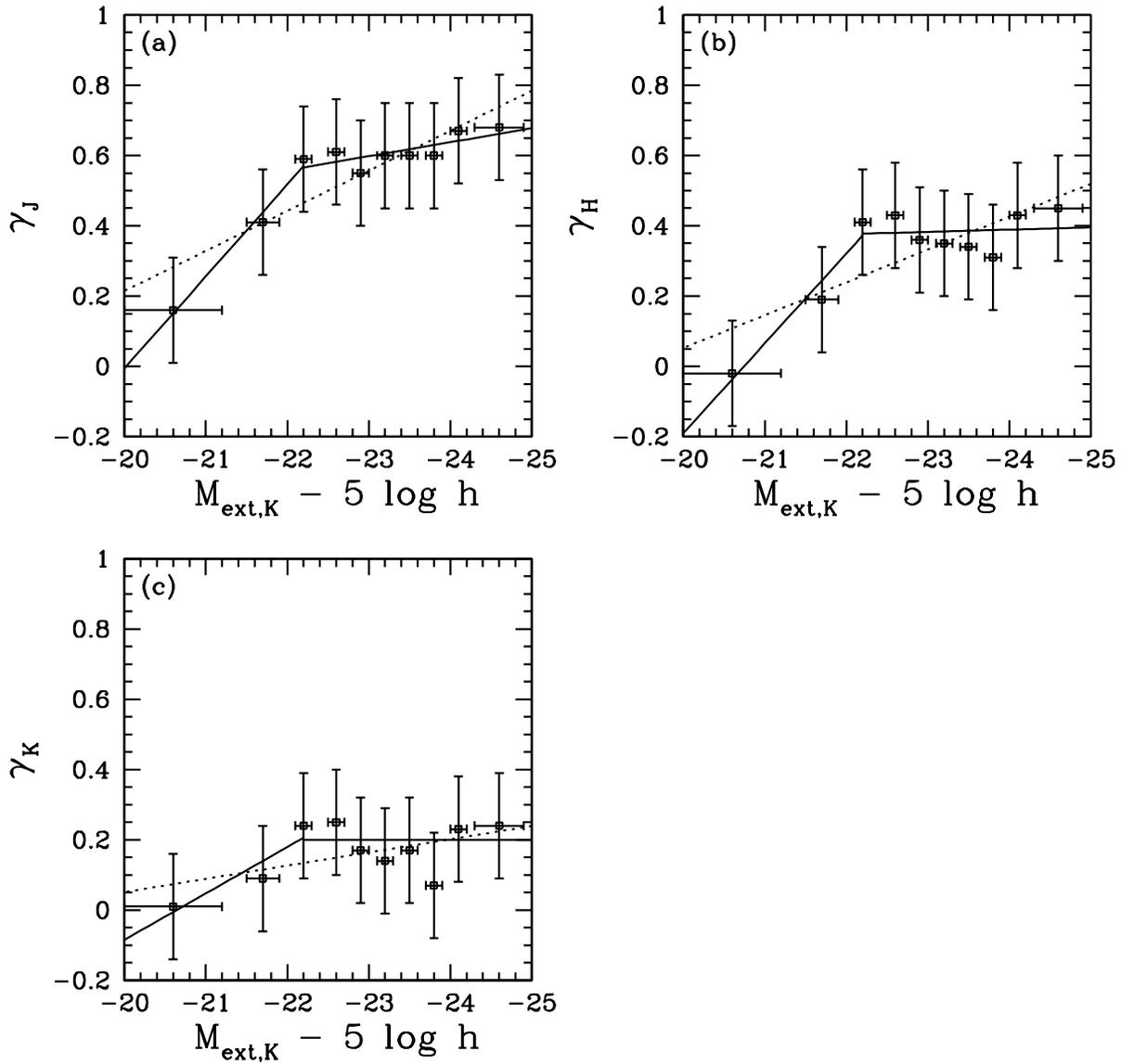}
\caption{Values of $\gamma$ as a function of luminosity in (a) J-band, (b) H-band and (c) K-band derived using the modified Holmberg test with $\delta$ as given by the truncated cubic fit to the points in Figure \ref{del-lum}. The dotted line shows a straight line fit to the points, the solid line  our adopted luminosity dependence.}
\label{gam-lum}
\end{figure*}

The extent of the luminosity dependence of $\gamma$ depends critically on the value $\delta$ used to correct radii to the face-on values. In the modified Holmberg test, an over-estimate of $\delta$ leads to an under-estimate of $\gamma$ (and vice-versa). Since $\delta$ clearly decreases with luminosity, using a constant value of $\delta$ for all galaxies thus leads to an over-estimate of $\gamma$ at low luminosities and an under-estimate at high luminosities and suppresses the expected luminosity dependence. 
The cubic fit in Figure \ref{del-lum} follows luminosity dependence of $\delta$ (under the assumption of transparency at \ri) and should be used here. The values of $\gamma$ derived by using this $\delta$ are shown in Figure \ref{gam-lum} as a function of luminosity. We find that over much of the range of luminosities the value of $\gamma$ is roughly constant in each band. In support of this we observe that the slope of the reddening does not change much with luminosity. It is clear that at low luminosities the extinction in these bands does decrease significantly and (unlike as found for galaxies in I-band by Giovanelli \etal ~1995) does fall off to zero for galaxies less luminous than $\MK=-20$, -20.7 and -20.9 in J, H and K\s ~respectively. The solid line in Figure \ref{gam-lum} shows our adopted luminosity dependence for $\gamma$, which is broken into two regions. The relation is $\gamma_J = -0.26 x + 0.77$,  $\gamma_H = -0.26 x + 0.58$, and  $\gamma_K = -0.13 x + 0.31$, for $x > 0.8$, and $\gamma_J = -0.04x + 0.60$, $\gamma_H = -0.01 x + 0.38$, and  $\gamma_K = 0.20$, for $x < 0.8$ (where $x=\MK + 23$). It should be noted that a straight line can fit within the errors on the points (dotted line in Figure \ref{gam-lum}), which are $\pm 0.15$ in $\gamma$. 

If the assumption of transparency at \ri ~is not valid then we are over-estimating $\delta$ and thus under-estimating $\gamma$. High luminosity galaxies have more extinction, but \ri$/r_d$ is also larger in these galaxies which could mean that they are more likely to be transparent at \ri ~than low luminosity galaxies. What bias the assumption of transparency might introduce into this determinations of the luminosity dependence of $\gamma$ is not clear.

\section{Conclusions}

Two samples of spiral galaxies are constructed, both of which have photometry from 2MASS and redshifts from the AGC. For the smaller sample (which has over 3000 galaxies), I-band photometry is also available. We apply a variety of statistical tests to look for disk inclination dependences in these galaxies. The results of these tests can be highly sensitive to the selection function of the sample and some interdependencies exist, but taken together they provide reliable information on the average internal extinction in the galaxies. We parameterize internal extinction via Eqn 7 and change of the isophotal radius via Eqn 10; these results are summarized in Tables \ref{results} and \ref{results2}, where the numbers between brackets indicate the error on the last digit of the parameter value.

\begin{deluxetable}{cccc}
\tabletypesize{\small}
\tablecaption{Summary of results for $\delta$ in $r_{21}/r_{21}^{\circ} = 1 + \delta \log(a/b)$. \label{results}}
\tablewidth{0pt}
\tablehead{
\colhead{Method} &  \colhead{J-band} & \colhead{H-band} & \colhead{K\s-band}}
\startdata
For transparent exponential disks & 0.9(3) & 0.7(2)& 0.7(2) \\
& & & \\
Statistical test & & & \\
lower limit &  0.6(1) & 0.4(1) & 0.2(1) \\
upper limit & 1.2(1) & 0.8(1) & 0.7(1)\\
& & & \\
Adopted & 0.9(1) & 0.7(1) & 0.7(1) \\
& & & \\
Luminosity dependence & $0.87 + 0.11 x $ & $0.70 + 0.07 x$ & $0.67 + 0.10 x$  \\
($x = \MK  + 23$) & $+ 0.05 x^2 + 0.007 x^3$ & $ + 0.05 x^2 + 0.009 x^3$ & $+ 0.05 x^2 - 0.004 x^3$ \\
\enddata
\end{deluxetable}

\begin{deluxetable}{cccc}
\tabletypesize{\small}
\tablecaption{Summary of results for $\gamma$ in $\Delta M = \gamma \log (a/b)$
\label{results2}}
\tablewidth{0pt}
\tablehead{
\colhead{Method} &  \colhead{J-band} & \colhead{H-band} & \colhead{K\s-band}}
\startdata
Reddening test & & & \\
All SFI2 & 0.3(1)+$\gamma_{\rm K}$ & 0.1(1)+$\gamma_{\rm K}$ & -- \\
& 0.2(1)+$\gamma_{\rm H}$& & \\
AGC$z$ & & & \\
low inclination\tablenotemark{a}  & 0.15(2)+$\gamma_{\rm K}$ & 0.05(2)+$\gamma_{\rm K}$  & -- \\
& 0.10(2)+$\gamma_{\rm H}$& -- & -- \\
high inclination\tablenotemark{a} & 0.46(5)+ 0.08$/\log(a/b)$ & 0.15(5) + 0.025$/\log(a/b)$  & -- \\
& 0.31(5) + 0.05$/\log(a/b)$& -- & -- \\
Statistical test & & & \\
lower limit & 0.2(1) & 0.1(1) & 0.2(1)\\
upper limit & 1.2(1) & 0.8(1) & 0.7(1)\\
& & & \\
Holmberg test\tablenotemark{b} & & & \\
low inclination\tablenotemark{a} & 0.48(15)  & 0.39(15)  & 0.26(15)  \\
high inclination\tablenotemark{a} & $\geq 1.6 + 0.24/\log(a/b) $ & $\geq 1.3 + 0.20/\log(a/b)$  & $\geq 1.1 + 0.13/\log(a/b)$  \\
& & & \\
Adopted & & & \\
low inclination\tablenotemark{a} & 0.48(15) & 0.39(15) & 0.26(15) \\
high inclination\tablenotemark{a} & [1.6(2) + 0.24$/\log(a/b)$]  & [1.4(2)$ + 0.20/\log(a/b)$] & [1.1(2) + 0.13$/\log(a/b)$] \\
& & & \\
Luminosity dependence & & & \\
($x = \MK  + 23$) & & & \\
$x  > 0.8$ & $-0.26x + 0.77$ & $-0.26x + 0.58$ & $-0.13x + 0.31$\\
$x < 0.8$ &$-0.04x + 0.60$ & $-0.01x + 0.38$  & 0.20 \\
\enddata
\tablenotetext{a}{Limit between low and high inclinations is $\log (a/b) = 0.50$.}
\tablenotetext{b}{Assumes $\delta=$0.9, 0.7 and 0.7 in J, H and K\s ~respectively.}
\end{deluxetable}

From the trends of NIR color with inclination in the SFI2 sample, we derive a firm lower limit to the total extinction in J and H-band of $\gamma_{\rm J}=0.3+\gamma_{\rm K}$ mag and $\gamma_{\rm H} = 0.1 +\gamma_{\rm K}$ mag respectively.
The scatter in these trends is on the order of 0.1 mag, while the statistical error on determining the slope is 0.01 mag. Using the same argument, we derive a lower limit on the extinction in I-band of $0.9+\gamma_{\rm K}$ mag, which when compared with those derived by G94 and \citet{t98}, is consistent with very low extinction in K\s-band. Using the much larger AGC$z$ sample, we note that the simple linear relation given by Eqn 7 does not adequately describe the variation of the 2MASS colors with inclination, but an upturn at high inclinations takes place instead which we fit with a bi-linear law.

The radius $r_{21}$ is seen to be closer to the center of the disk in J, H and K\s ~($\sim 2$-3 scale lengths) than $r_{23.5}$ in I-band (at 3.5-4 scale lengths), and we work through a simple model to estimate at what inclination the disk might become opaque at that radius for a given central face-on opacity. 
Using samples defined according to different selection criteria, we explore the inclination dependence of \ri ~and compare the data with the dependence that would be expected for pure exponential disks which are transparent at \ri ~at all inclinations. {\em We find that complete transparency at \ri ~is inconsistent with the data of very nearly edge-on galaxies.} Upper and lower limits are estimated for $\delta$ (as defined by Eqn 10), and we adopt as most probable values $\delta_{\rm J} \simeq 0.9 \pm 0.1$, $\delta_{\rm H} = 0.7 \pm 0.1$ and $\delta_{\rm K} = 0.7 \pm 0.1$.

Tracking the flux dependence on inclination and using a modified version of the Holmberg test we investigate the amount of internal extinction in disks in the three bands. This analysis confirms that a simple value of $\gamma$ (as defined by Eqn 7) is not adequate to describe extinction for all values of $\log(a/b)$. 
Because of this we fit a bi-linear law fitting separately high and low inclination systems as separated by $\log(a/b)=0.5$. The value of $\gamma$ calculated from this test is sensitive to how the isophotal radii are corrected to the face-on value (\ie ~the value of $\delta$ that is used in Eqn 10) and can change by a significant amount. Using the adopted values of $\delta$ we find $\gamma_{\rm J} = 0.48 \pm 0.15$, $\gamma_{\rm H} = 0.39 \pm 0.15$, and $\gamma_{\rm K} = 0.26 \pm 0.15$ for $\log(a/b) < 0.5$. If $\delta$ is increased, $\gamma$ decreases and vice-versa. 

At the {\it higher inclinations} the modified Holmberg test is not appropriate as we expect a much smaller variation of \ri ~with inclination than is corrected for. This implies that the measured slope is only a lower limit to the value of $\gamma$ in Eqn 7. We expect that the limit will be closest to the actual value of $\gamma$ in K\s ~where the extinction is lowest and therefore adopt $\gamma_{\rm K} = 1.1 \pm 0.2 + 0.13/\log(a/b)$. Using information on reddening we then derive $\gamma_{\rm J} = 1.6 \pm 0.2 + 0.24/\log(a/b)$ and $\gamma_{\rm H} = 1.4 \pm 0.2 + 0.20/\log(a/b)$.

We explore a simple photometric model for extinction due to dust in spiral galaxies. In the model the dust is distributed exponentially both radially and in the $z$-direction. The scale length of the dust is assumed to be the same as the stars, while a ratio $\zeta$ is assumed between the scale heights. The ``best fit'' models to the data (all with $\zeta=0.5$) suggest a central face-on opacity, $\tau^\circ(0) = 1.1$ and intrinsic axial ratio, $q=0.14$ for J-band, $\tau^\circ(0) = 0.7$ and $q=0.14$ for H-band and $\tau^\circ(0) = 0.3$ and $q=0.15$ for K\s-band. This model explains well the upturn in the extinction at high inclinations, which becomes more obvious for lower values of $\tau^\circ(0)$.

 In Tables 4 and 5 we give estimates of our favored solutions for $\delta$ in Eqn 10 and $\gamma$ in Eqn 7 as compromises of derivations from different methods. They are listed in the row labelled ``Adopted''.

Finally, we explore the luminosity dependence of $\delta$ and $\gamma$.  
Under the assumption of transparency at \ri ~the value of $\delta$ decreases with luminosity. We use the modified Holmberg test 
to explore the luminosity dependence of $\gamma$ and find that the extinction drops off to zero below a luminosity of $\MK=-20$, -20.7 and -20.9 in J, H and K\s ~respectively. and is roughly constant at $\gamma=0.61$, 0.39 and 0.19 at the higher luminosities. There is marginal evidence of further increase at the very highest luminosities. The adopted luminosity dependences are also listed in Tables 4 and 5. 

We have explored the commonly held view that there is little internal extinction in external galaxies at NIR wavelengths, especially K\s-band. We have shown that except for the least luminous galaxies, there is evidence for internal extinction (which increases with luminosity), and that in the most inclined galaxies the extinction becomes significant.

\acknowledgments{
We are very grateful to Dr. Tom Jarrett for his help in accessing the 2MASS Full release Extended Source Catalog and for his reading and comments on the paper. This publication makes use of data products from the Two Micron All Sky Survey, which is a joint project of the University of Massachusetts and the Infrared Processing and Analysis Center/California Institute of Technology, funded by the National Aeronautics and Space Administration and the National Science Foundation. This work has been partially supported by the NSF Grant No. AST-0098526.
}

\clearpage
\onecolumn

\begin{thebibliography}{}
 
\bibitem[Calzetti (2001)]{cal01} Calzetti, D., 2001, \pasp, 113, 1449
\bibitem[Curtri \etal ~(2000)]{cutri} Cutri, R. M., \etal ~2000, Explanatory Supplement to the 2MASS Second Incremental Data Release (Pasadena: Caltech)
\bibitem[Disney \& Burstein (1995)]{opacitybook} Disney, J. I., Burstein, D. Eds. 1995, The Opacity of Spiral Disks (Dordrecht: Kluwer)
\bibitem[Disney \etal ~(1989)]{disney} Disney, M., Davis, J. \& Phillips, S. 1989, \mnras, 239, 939
\bibitem[Evans (1992)]{evans92} Evans, Rh. 1992, Ph.D. thesis, University of Cardiff
\bibitem[Giovanelli \etal ~(1994)]{g94} Giovanelli, R., Haynes, M. P., Salzer, J. J., Wegner, G., da Costa, L.N., \& Freudling, W. 1994, \aj, 107, 2036 (G94)
\bibitem[Giovanelli \etal ~(1995)]{g95} Giovanelli, R., Haynes, M. P., Salzer, J. J., Wegner, G., da Costa, L.N., \& Freudling, W. 1995, \aj, 110, 1059
\bibitem[Giovanelli \etal ~(1997)]{g97} Giovanelli, R., Haynes, M., Herter, T., Wegner, G., Salzer, J. J., da Costa, L. N., \& Freudling, W. 1997, \aj, 113, 22 
\bibitem[Graham (2001)]{graham01} Graham, A. W. 2001 \mnras, 326, 543 
\bibitem[Haynes \etal ~(1999a)]{h99a} Haynes, M. P., Giovanelli, R., Salzer, J. J., Wegner, G., Freudling, W., da Costa, L. N., Herter, T., \& Vogt, N. P. 1999a, \aj, 117, 1668
\bibitem[Haynes \etal ~(1999b)]{h99b} Haynes, M. P., Giovanelli, R., Chamaraux, P., da Costa, L. N., Freudling, W., Salzer, J. J., Wegner, G. 1999b, \aj, 117, 2039
\bibitem[Holmberg (1958)]{holm} Holmberg, E. 1958, Medd. Lunds Astr Obs. Ser. 2, No. 136
\bibitem[Holmberg (1975)]{holm2} Holmberg E. 1975, in Stars and Stellars Systems, Vol. IX, A. Sandage, M. Sandage, J. Kristian eds. (University of Chicago, Chicago).
\bibitem[Huizinga (1994)]{Huzthesis} Huizinga, J. E. 1994, Ph.D. thesis, University of Groningen
\bibitem[Jarrett \etal ~(2003)]{J03} Jarrett, T. H., Chester, T., Cutri, R., Schneider, S., \& Huchra, J. P., 2003, \aj, 125, 525.
\bibitem[Jarrett \etal ~(2000)]{J00} Jarrett, T. H., Chester, T., Cutri, R., Schneider, S., Skrutskie, M., \& Huchra, J.P. 2000, \aj, 119, 2498
\bibitem[de Jong (1996)]{deJongII} de Jong, R. S. 1996, \aaps, 118, 557
\bibitem[Kochanek \etal ~(2001)]{kochanek01} Kochanek, C. S., Pahre, M. A., Falco, E. E., Huchra, J. P., Mader, J., Jarrett, T. H., Chester, T., Cutri, R., Schneider, S. E. 2001, \apj, 560, 566
\bibitem[Kron (1980)]{kron} Kron, R. G. 1980, \apjs, 43, 305  
\bibitem[Kuchinski \etal ~(1998)]{kuchinski98} Kuchinski, L. E., Terndrup, D. M., Gordon, K. D. \& Witt, A. N. 1998, \aj, 115 1438
\bibitem[Moriondo \etal ~(1998)]{moriondo} Moriondo, G., Giovanelli, R., \& Haynes, M. P. 1998, \aap, 338, 795
\bibitem[Poggianti (1997)]{pog} Poggianti, B. M. 1997, \aaps, 122, 399
\bibitem[Schlegel \etal ~(1998)]{dirbe} Schlegel, D. J., Finkbeiner, D. P., \& Davis, M. 1998, \aj, 500, 525
\bibitem[Tully \etal ~(1998)]{t98} Tully, R. B., Pierce, M. J., Huang, J-S., Saunders, W., Verheijen, M. A. W., Witchalls, P. L. 1998, \aj, 115, 2264
\bibitem[Tully \& Fisher (1977)]{tf} Tully, R. B. \& Fisher, J. R. 1977, \aap, 54, 661
\bibitem[Xilouris \etal ~(1999)]{xi99} Xilouris, E. M., Byun, Y. I., Kyafis, N. D., Paleologou, E. V., \& Papamastorakis, J. 1999, \aap, 344, 868
\end{thebibliography}
\end{document}